\documentclass[%
 aip,
 jmp,%
 amsmath,amssymb,
 reprint,%
]{revtex4-1}

\usepackage{graphicx}
\usepackage{dcolumn}
\usepackage{bm}
\usepackage[mathlines]{lineno}
\usepackage{amssymb,amsfonts,amsmath}
\usepackage{color}
\usepackage{amsmath,amssymb,amscd,ulem}
\usepackage{comment}
\usepackage{framed}
\usepackage{here}

\newcommand{\ProbP}{\mathbb{P}}
\newcommand{\ProbQ}{\mathbb{Q}}

\newcommand{\ProbT}{\mathbb{T}}

\newcommand{\LF}{K}
\newcommand{\lf}{k}
\newcommand{\defeq}{:=}
\newcommand{\const}{\mathrm{const.}}
\newcommand{\CMF}{\Psi}

\newcommand{\MI}{\mathcal{I}}
\newcommand{\average}[1]{\left< #1 \right>}

\newcommand{\KLD}{\mathcal{D}}
\newcommand{\KL}[2]{\KLD\Big[#1\Big|\Big|#2\Big]}

 \DeclareMathAlphabet{\mathpzc}{OT1}{pzc}{m}{it}\newcommand{\xpzc}{\mathpzc{X}}
\newcommand{\ypzc}{\mathpzc{Y}}
\newcommand{\zpzc}{\mathpzc{Z}}

\newcommand{\popN}{\mathcal{N}}
\newcommand{\com}[1]{#1}
\newcommand{\eqnref}[1]{eq. (\ref{#1})}
\newcommand{\Eqnref}[1]{Equation (\ref{#1})}

\newcommand{\setx}{\mathfrak{S}_{x}}
\newcommand{\sety}{\mathfrak{S}_{y}}
\newcommand{\setz}{\mathfrak{S}_{z}}
\newcommand{\elx}{\mathfrak{s}^{x}}
\newcommand{\ely}{\mathfrak{s}^{y}}
\newcommand{\elz}{\mathfrak{s}^{z}}
\newcommand{\setX}{\mathfrak{S}_{\xpzc}}
\newcommand{\setY}{\mathfrak{S}_{\ypzc}}
\newcommand{\setZ}{\mathfrak{S}_{\zpzc}}

\newcommand{\loss}{\mathrm{loss}}

\newcommand{\cs}{s}

\begin{document}

\preprint{AIP/123-QED}

\title[Stochastic and Information-thermodynamic Structure in Adaptation]{Stochastic and Information-thermodynamic Structures of Population Dynamics in Fluctuating Environment}

\author{Tetsuya J. Kobayashi}
 \email{tetsuya@mail.crmind.net}
 \homepage{http://research.crmind.net}
 \affiliation{Institute of Industrial Science, the University of Tokyo.}
 \affiliation{PRESTO, JST.}
\author{Yuki Sughiyama}%
 \affiliation{Institute of Industrial Science, the University of Tokyo.}

\date{\today}

\begin{abstract}
Adaptation in a fluctuating environment is a process of fueling environmental information to gain fitness.
Living systems have gradually developed strategies for adaptation from random and passive diversification of the phenotype to more proactive decision making, in which environmental information is sensed and exploited more actively and effectively. 
Understanding the fundamental relation between fitness and information is therefore crucial to clarify the limits and universal properties of adaptation.
In this work, we elucidate the underlying stochastic and information-thermodynamic structure in this process, by deriving  causal fluctuation relations (FRs) of fitness and information. 
Combined with a duality between phenotypic and environmental dynamics, the FRs reveal the limit of fitness gain, the relation of time reversibility with the achievability of the limit, and the possibility and condition for gaining excess fitness due to environmental fluctuation.
The loss of fitness due to causal constraints and the limited capacity of real organisms is shown to be the difference between time-forward and time-backward path probabilities of phenotypic and environmental dynamics.
Furthermore, the FRs generalize the concept of evolutionary stable state (ESS) for fluctuating environment by giving the probability that the optimal strategy on average can be invaded by a suboptimal one owing to rare environmental fluctuation. 
These results clarify the information thermodynamic structures in adaptation and evolution.
\end{abstract}

\pacs{Valid PACS appear here}
\keywords{Fluctuation theorem; Evolution; Decision making; Bet-hedging;Fitness; Variational structure;}
\maketitle


\section{Introduction}
\subsection{Adaptation in fluctuating environment}
Adaptation is fundamental to all organisms for their survival and evolutionary success in a changing environment.
In the course of evolution, living systems have gradually attained and developed more active and efficient strategies for adaptation, which generally accompany more effective use of environmental information.
Understanding how the use of information is linked to the efficiency of adaptation is crucial to clarify the fundamental limits and universal properties of biological adaptations\cite{Chevin:2010cw,Frank:2011cq}. 

The most primitive strategy for adaptation is to randomly generate genetic and phenotypic heterogeneity in a population \cite{Slatkin:1974it,Philippi:1989bm,Balaban:2004bq,Wakamoto:2013hb}.
Provided that a sufficiently large heterogeneity is constantly generated in the population,  a fraction of organisms can, by chance, have the types adaptive to  the upcoming environmental state and circumvent extinction of the population at the cost of others with non-adaptive types\cite{Cohen:1966ub,Cohen:1971wm}.
Such a strategy is known as bet-hedging or phenotypic diversification and works even if the organisms are completely blind to the environment, without any \textit{a priori} knowledge of its dynamics.
The bet-hedging is a passive and \textit{a posteriori} adaptation in the sense that the adaptation is achieved extrinsically by and after the impact of environmental selection\cite{Kobayashi:2015eca}.
The evolutionary advantage of the bet-hedging strategy is demonstrated by the persistence of bacteria, pathogens, and cancer cells to antibiotic or anticancer drug treatments\cite{Balaban:2004bq,Dhar:2007csa,Beaumont:2009gxa,Meacham:2013ge,Wakamoto:2013hb}.
The gain of fitness by bet-hedging can be optimized if the population evolves to generate  an appropriate pattern of heterogeneity by learning the environmental statistics\cite{Xue:2016fy}.
Nevertheless, the gain of fitness by bet-hedging is fundamentally limited because of the passive and \textit{a posteriori} nature of the strategy, in which the individual organisms have no access to the microscopic information of which environmental states will actually be realized.

With any access to such information, the loss can be avoided further by decision making: directly sensing the current environmental state, predicting the upcoming state, and switching into the phenotypic state that is adaptive to that state \cite{Perkins:2009cg,BenJacob:2010ii,Kobayashi:2012ji,Brennan:2012cj}. 
The strategy of adaptation via sensing is active and \textit{a priori} in a sense that adaptation is intrinsically achieved by  the predictive actions of the organisms\cite{Kobayashi:2015eca}.
In biologically relevant situations, both passive and active aspects of adaptation are intermingled because perfect sensing and prediction of environment are impossible with the limited capacities of biological systems.

\subsection{Notions of information and analogy with physics in biological adaptation}
At an analogical level, the problem of the fundamental law and the limits of adaptation and evolution shares several aspects with physics, especially with thermodynamics, which drove the long-lasting attempts to establish the thermodynamics of biological adaptation and evolution \com{\cite{Lotka:1922ux,Lotka:1922wr,Iwasa:1988ux,deVladar:2011kz,Frank:2012fq,Qian:2014ha,Mustonen:2010ig}.}
Among other areas, the fundamental limit of fitness in a changing environment and the value of environmental information have been a major focus in evolutionary biology \cite{Levins:1965a,Levins:1968tc,Cohen:1966ub,Chevin:2010cw,Frank:2011cq,Rivoire:2015if}. 
Haccou and Iwasa may be the first who linked, albeit implicitly, environmental information with the gain of fitness in a stochastic environment\cite{Haccou:1995tf}.
Bergstrom and Lachmann pursued the fitness value of information by directly incorporating mutual information\cite{Bergstrom:2004um,DonaldsonMatasci:2010ie,Cover:2012ub}.
Others also pointed out some quantitative relations between fitness and information measures such as relative entropy and Jeffreys' divergence\cite{Kussell:2005dg,Frank:2012kq,Pugatch:2013va}.
More recently, Rivoire and Liebler conducted a comprehensive analysis  by employing an analogy between bet-hedging of organisms and horse race gambling \cite{Rivoire:2011fy}, the link of which to information theory was revealed in the seminal work by Kelly in 1956 \cite{Kelly:1956dy}.
However, all previous works either imposed certain restrictions on their models to derive the information-theoretic measures of fitness value \cite{Kussell:2005dg,Rivoire:2011fy,Rivoire:2014kt} or had to introduce phenomenological measures  for the value of information to accommodate more general situations\cite{Rivoire:2011fy,Rivoire:2014kt},
 because they lacked an appropriate method to handle the mixture of the passive and active aspects in adaptation.

We recently resolved this problem \cite{Kobayashi:2015eca} by combining a path integral formulation of population dynamics \cite{Leibler:2010jx,Bianconi:2012kz,Wakamoto:2012hx,Sughiyama:2015cf}, a retrospective characterization of the selected population \cite{Baake:2006ek,Wakamoto:2012hx,Lambert:2015bk}, and a variational structure in population dynamics \cite{Demetrius:2014et,Sughiyama:2015cf}.
The results we obtained generalized the limits of fitness gain by sensing and revealed that the gain satisfies fluctuation relations (FRs) that fundamentally constrain not only its average but also its fluctuation.
These relations, \com{alongside a previous work in the line of neutral theory \cite{Mustonen:2010ig}}, imply that fitness in the fluctuation environment shares, at least mathematically, similar structures to those of  stochastic and information thermodynamics\cite{Seifert:2012es,Sagawa:2012wi}.
In our FRs, the fluctuation of fitness of a given population is evaluated by the difference from the fitness that achieves the maximum average fitness over all possible phenotypic histories of organisms.
Conceptually, this means that we postulate a Darwinian demon, an imaginary organism, that can exhibit any type of behavior without imposing any constraint not only on biological capacity but also on the causality of dynamics.
The FRs characterize the loss of fitness of a realistic organism from such an idealized organism.
Thus, understanding the properties of the Darwinian demon and the deviation from it by a realistic organism are central to a deeper understanding of the behavior of populations in a changing environment.
However, the implicit definition of the demon as the maximizer of the average fitness hampers the explicit characterization of the demon and obscures the formal link to stochastic thermodynamics , in which a variational characterization is not common\cite{Seifert:2012es,Sagawa:2012wi}.
More practically, without an explicit characterization, we are unable to simulate possible behaviors of the demon even numerically.

\subsection{Outline of main results}
In this paper, we resolve these problems by deriving FRs of fitness without using the variational approach.
To this end, we first formulate and generalize the problem of adaptation in a changing environment so that individual organisms can change not only their strategy of switching phenotypic states but also the strategy of allocating metabolic resources to each phenotypic state (Sec. \ref{sec:modeling}).
Combined with the path integral formulation of population dynamics, this generalization enables us to obtain a decomposition of fitness with a combination of time-forward (chronological) and time-backward (retrospective) path probabilities (Sec. \ref{sec:PathFormDecomp}).
The decomposition naturally spells out an explicit representation of the upper bound of the average fitness,  which was implicitly defined in our previous work\cite{Kobayashi:2015eca}.

For the bet-hedging problem without a sensing environment (Sec. \ref{sec:bethedging}), the decomposition directly leads to FRs of the fitness loss, which has a very similar form as the FRs of entropy production in stochastic thermodynamics\cite{Seifert:2012es}.
After numerically verifying the derived FRs (Sec. \ref{ssec:VerFRbf}), 
we investigate the biological meanings and achievability of the FRs  (Sec. \ref{sec:BMFRbh}).
The average FR is related to the evolutionary stable state (ESS) , under which the strategy with maximal average fitness cannot be invaded by any other strategies.
The detailed and integral FRs generalize the ESS by giving the probability that a suboptimal strategy outperforms the optimal one within a finite time interval owing to rare fluctuation of the environment\cite{King:2007fi}.
By using a dualistic relation between phenotypic and environmental dynamics,  the detailed FR is shown to be represented as the ratio of the path probability of the actual environment and that of the conjugate environment under which a given strategy of the organisms becomes optimal.
The duality also clarifies that the average loss of fitness is directly related to the imperfectness of the adaptive behavior of the organisms, originating both from physical constraints and from the suboptimality of the behaviors.

The introduction of a sensing signal extends the FRs to accommodate the mutual information between the environment and the signal as the gain of fitness by sensing, in the same manner as mutual information bounds the negative gain of entropy production in information thermodynamics\cite{Sagawa:2012wi} (Sec. \ref{sec:FRs}).
Although the extended FRs cover very general situations,  the FRs are not tight and therefore the mutual information overestimate the value of fitness by sensing.
By explicitly assuming a causal relation between the environment and the signal, 
the FRs are further modified to involve the directed information as a tighter bound of fitness gain (Sec. \ref{sec:CFRs}).
This modification clarifies how the loss of fitness from the upper bound is related to the causality,  the inaccessibility to perfect information of the environment, and the imperfect implementation of information processing.
Finally, three quantities are introduced to account for the fitness loss of inappropriate sensing and the imperfectness of metabolic allocation and phenotypic switching strategies  in general situations (Sec. \ref{sec:MMIs}).
The summary and future directions are described in Discussion (Sec. \ref{sec:Sum}).


\section{Modeling adaptation of population in changing environment}\label{sec:modeling}
Let $x_{t} \in \setx $, $y_{t} \in \sety$, and $z_{t} \in \setz$ be the phenotype of a living organism, the state of the environment, and the state of the sensing signal at time $t$, respectively (Fig. 1 (A)).
For simplicity, possible phenotypic, environmental, and signal states are assume to be discrete as in references \cite{DonaldsonMatasci:2010ie,Rivoire:2011fy,Pugatch:2013va,Kobayashi:2015eca}. 
The paths (histories) of the states up to time $t$ are defined as
$\xpzc_{t} \defeq \{x_{\tau}|\tau \in [0, t]\} \in \setX \defeq \setx^{\times (t+1)}$, $\ypzc_{t} \defeq \{y_{\tau}|\tau \in [0, t]\}\in \setY \defeq \sety^{\times (t+1)}$, and $\zpzc_{t} \defeq \{z_{\tau}|\tau \in [0, t]\}\in \setZ \defeq \setz^{\times (t+1)}$, respectively.
Time is also treated as discrete in this work.

\subsection{Modeling phenotype switching}
The phenotype of an organism, in general, switches stochastically over time, depending on its past phenotypic state and the sensed signal (Fig. 1 (A) and (B)).
The switching dynamics is modeled, for example, by a Markov transition probability $\ProbT_{F}(x_{t+1}|x_{t},z_{t+1})$, which satisfies $\sum_{x_{t+1}}\ProbT_{F}(x_{t+1}|x_{t},z_{t+1})=1$  for all $x_{t}$ and $z_{t+1}$.
Although we mainly focus on the Markov switching, our result can be extended for causal switching $\ProbT_{F}(x_{t+1}|\xpzc_{t},\zpzc_{t+1})$ in which the next phenotypic state $x_{t}$ depends on both past phenotypic and signal histories $\xpzc_{t}$ and $\zpzc_{t+1}$.

\subsection{Modeling metabolic resource allocation}
Next, we model the strategy of metabolic resource allocation of the organisms. 
Each organism is assumed to duplicate asexually to produce $e^{\lf}$ daughter organisms on average. 
For each state of environment $y$, the organisms have a maximum replication rate, $e^{\lf_{max}(y)}$, that can be achieved only when the organism allocates its all metabolic resources for adapting only to that environmental state. 
Because the environment changes, however, the organism usually distributes its resources for different environmental conditions\cite{Giordano:2016gt}. 

To represent this situation, we introduce a conditional probability $\ProbT_{\LF}(y|x)$ that quantitatively represents the fraction of resources allocated to the environmental state $y$ in a phenotype $x$.
An instantaneous replication rate $\lf(x,y)$ of the phenotype $x$ under the environmental state $y$ is then assumed to be represented as 
\begin{align}
e^{\lf(x,y)}=e^{\lf_{max}(y)}\ProbT_{\LF}(y|x), \label{eq:ek}
\end{align}
 (Fig. 1 (C)).
Note that such a decomposition of $e^{\lf(x,y)}$ can date back to Haccou and Iwasa\cite{Haccou:1995tf}, at least.
This relation between the resource allocation strategy and the replication rate may appear to be restrictive because of the linear relation between the allocation strategy $\ProbT_{\LF}(y|x)$ and the actual replication rate $e^{\lf(x,y)}$.
Nevertheless, this decomposition of $\lf(x,y)$ is general enough because, for a given $\lf(x,y)$, we can find a pair of $e^{\lf_{max}(y)}$ and $\ProbT_{\LF}(y|x)$ as long as the possible phenotypic states are fewer than those of the environment (refer to Appendix \ref{ap:decomp}).
Such a situation is biologically plausible because the environment is usually more complex than the phenotype of an organism.
Moreover, our setting includes the special situation that has been intensively investigated in previous works\cite{Kussell:2005dg,Rivoire:2011fy}.
For example, when the numbers of phenotypic and environmental states are equal as $\# \setx= \# \sety$,$\ProbT_{\LF}(y|x)=\delta_{x,y}$ corresponds to Kelly's horse race gambling \cite{Kelly:1956dy,Rivoire:2011fy}, in which each phenotypic state allocates all metabolic resources to a certain environmental state and, as a result, can survive and grow only when the phenotypic state matches the realized environmental state.

\subsection{Modeling sensing processes}
We finally model the sensing process. 
We consider the case that organisms can obtain a sensing signal $z(t)$, which correlates with the environment $y(t)$.
The sensed signal $z(t)$ is assumed to be common to all the organisms in the population (Fig. 1 (A)).
A biologically relevant situation is that $z(t)$ is a vector of concentrations of extracellular signaling molecules that cannot be consumed as metabolites but correlate with the available metabolites. 
Another situation is that $z(t)$ is a subset of $y(t)$ to which the organisms have sensors.
In either situation, the sensing noise should be negligibly small because all the organisms receive the same sensing signal $z(t)$.
Even though sensing of a common signal cannot cover all biologically realistic situations such as individual sensing with noisy receptors\cite{Perkins:2009cg,BenJacob:2010ii,Kobayashi:2010vo,Kobayashi:2012ji,Brennan:2012cj,Barato:2014df}, the common signal has been investigated in various works \cite{Haccou:1995tf,Rivoire:2011fy,Kobayashi:2015eca}.
In this work, we mainly focus on the common sensing problem and touch on the individual sensing in Discussion.



\begin{figure}
\includegraphics[width=\linewidth]{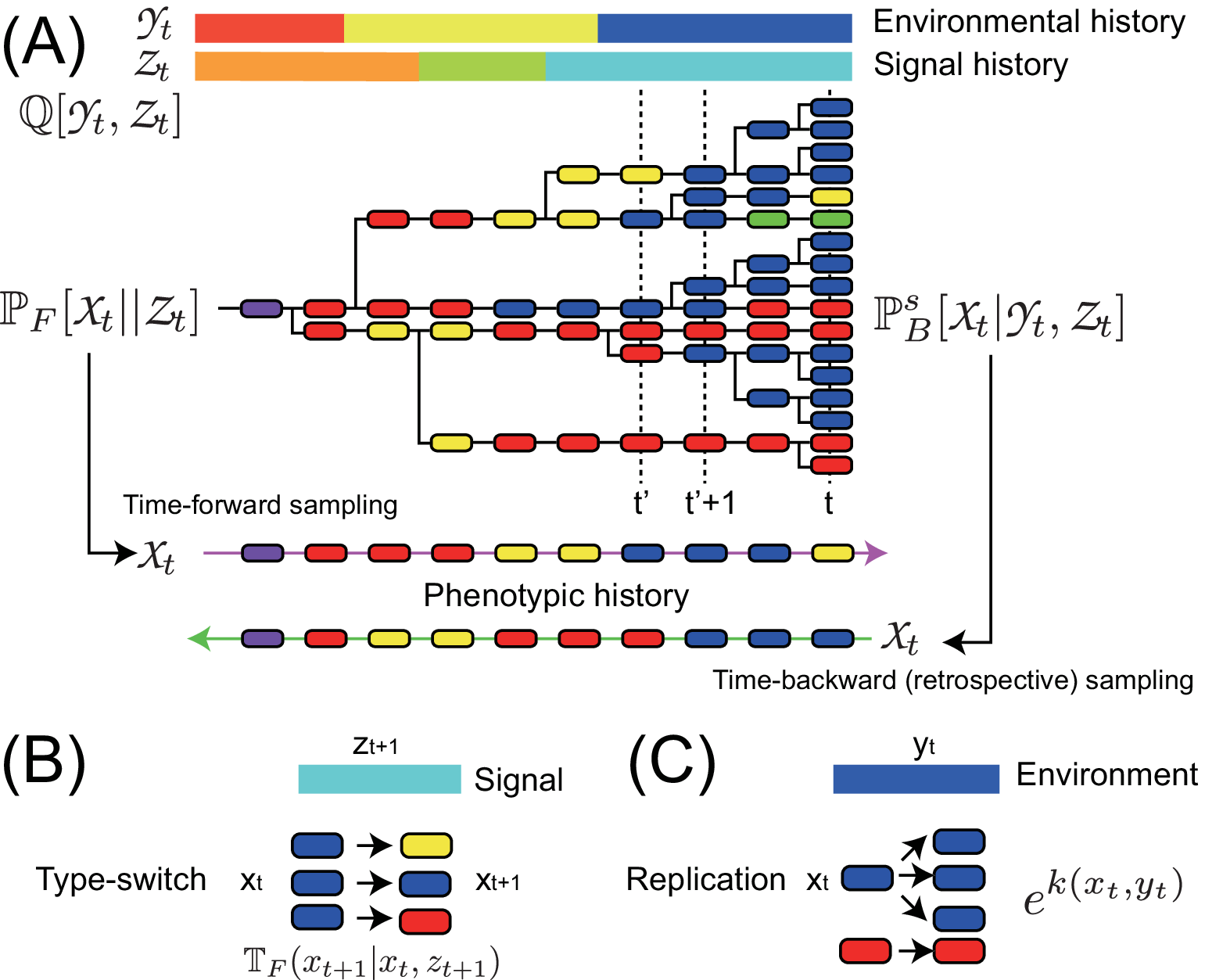}
\caption{\label{fig1} Schematic diagrams for population dynamics of organisms with multiple types and sensing in a changing environment. 
(A) A tree representation of a growing population in a changing environment. Colors indicate different phenotypic states, environmental states, and signal over time. By tracking the phenotypic history from $t=0$ to $t$ in a time-forward manner, we can \com{effectively obtain a sample $\xpzc_{t}$ with probability} $\ProbP_{F}[\xpzc_{t}||\zpzc_{t}]$. 
By tracking the phenotypic history retrospectively from $t$ to $t=0$ in a time-backward manner, we can obtain a sample \com{$\xpzc_{t}$ with probability} $\ProbP_{B}^{\cs}[\xpzc_{t}|\ypzc_{t}, \zpzc_{t}]$.  
\com{Note that we assume that the size of the population is sufficiently large when we introduce the population dynamics of the organisms \eqnref{eq:popdyn}.}
(B) Phenotypic switching from $x_{t}$ to $x_{t+1}$ in response to environmental signal $z_{t+1}$. The switching probability is represented by $\ProbT_{F}(x_{t+1}|x_{t},z_{t+1})$.
(C) Replication of organisms with phenotypic state $x_{t}$ under environmental state $y_{t}$.
Each organism generates $e^{\lf(x_{t},y_{t})}$ descendants on average.
}
\end{figure}

\subsection{Population dynamics of organisms}
By combining the phenotype switching, the metabolic allocation, and the sensing strategies, we can explicitly derive the dynamics of the population of the organisms.
Because both environmental and sensing histories are external factors of the organisms, the population dynamics of the organisms is described for a given pair of environmental and signaling histories, $\ypzc_{t}$ and $\zpzc_{t}$. 
Let $\popN_{t}^{\ypzc,\zpzc}(x_{t}) \in \mathbb{R}_{\ge 0}$ be the number of organisms whose phenotypic state is $x_{t}$ at time $t$ under the realization of environmental and signal histories $\ypzc_{t}$ and $\zpzc_{t}$.
\com{The population size of the organisms is assumed to be sufficiently large so that $\popN_{t}^{\ypzc,\zpzc}(x_{t})$ can be well approximated as a continuous variable. }
\com{Because the large population size enable us to effectively ignore the demographic fluctuation due to finite number of the organisms, 
we can obtain the update rule\cite{Kobayashi:2015eca} of $\popN_{t}^{\ypzc,\zpzc}(x_{t})$ as }
\begin{align}
\popN_{t+1}^{\ypzc,\zpzc}(x_{t+1})=e^{\lf(x_{t+1},y_{t+1})}\sum_{x_{t}\in \setx}\ProbT_{F}(x_{t+1}|x_{t},z_{t+1})\popN_{t}^{\ypzc,\zpzc}(x_{t}). \label{eq:popdyn}
\end{align}
\com{If we need to work on a population, the size of which is not sufficiently large, modeling of the population e.g., by  a branching process is required.}
The statistical properties of the environmental and signal histories are generally characterized by a joint path probability $\ProbQ[\ypzc_{t},\zpzc_{t}]$.

\section{Path-wise formulation and fitness decomposition}\label{sec:PathFormDecomp}
By using $\popN_{t}^{\ypzc,\zpzc}(x_{t})$ in the previous section, we define the fitness of the population and derive its path integral formulation.
The cumulative fitness $\CMF^{s}[\ypzc_{t},\zpzc_{t}]$ of the population at $t$ under environmental and signal histories $\ypzc_{t}$ and $\zpzc_{t}$ is defined by the exponential expansion of the total population size as follows:
\begin{align}
\CMF^{\cs}[\ypzc_{t},\zpzc_{t}] \defeq \ln\frac{\sum_{x_{t}}\popN_{t}^{\ypzc,\zpzc}(x_{t})}{\sum_{x_{0}}\popN_{0}^{\ypzc,\zpzc}(x_{0})}
=\ln\frac{\popN_{t}^{\ypzc,\zpzc}}{\popN_{0}^{\ypzc,\zpzc}}.
\end{align}
We here define $\popN^{\ypzc,\zpzc}_{t}\defeq \sum_{x_{t}}\popN^{\ypzc,\zpzc}_{t}(x_{t})$.
The ensemble average of the cumulative fitness for different realizations of the environmental and signal histories is represented as
\begin{align}
\average{\CMF^{\cs}_{t}}_{\ProbQ} \defeq \average{\CMF^{\cs}[\ypzc_{t},\zpzc_{t}]}_{\ProbQ[\ypzc_{t},\zpzc_{t}]}.
\end{align} 

\subsection{Path-wise and retrospective formulation}
As derived in \cite{Leibler:2010jx,Sughiyama:2015cf,Kobayashi:2015eca}, the cumulative fitness at time $t$ can be represented with a path-wise (path integral) formulation. 
Let us first define the time-forward path probability of the phenotype switching as 
\begin{align}
\ProbP_{F}[\xpzc_{t}||\zpzc_{t}] &\defeq \left[\prod_{\tau=0}^{t-1}\ProbT_{F}(x_{\tau+1}|x_{\tau},z_{\tau+1})\right]p(x_{0}),
\end{align}
where $p(x_{0}) \defeq \popN_{0}^{\ypzc,\zpzc}(x_{0})/\sum_{x_{0}}\popN_{0}^{\ypzc,\zpzc}(x_{0})$.
We here use Kramer's causal conditioning $||$ rather than the usual conditioning $|$ in order to indicate that the path probability $\ProbP_{F}[\xpzc_{t}||\zpzc_{t}] $ is causally generated by the Markov transition matrix $\ProbT_{F}(x_{t+1}|x_{t},z_{t+1})$, which depends only on the past phenotypic state $x_{t}$ and the signal $z_{t+1}$\cite{Kramer:1998ufa,Permuter:2011jra}.
The single bar $|$ is also used for the normal conditioning of a path probability that does not necessarily satisfy the causal relation between conditioning and conditioned histories.
We also define the path-wise (historical) fitness of a phenotypic history $\xpzc_{t}$ under an environmental history $\ypzc_{t}$ as 
\begin{align}
\LF[\xpzc_{t},\ypzc_{t}]\defeq \sum_{\tau=0}^{t-1}\lf(x_{\tau+1}, y_{\tau+1}).
\end{align} 
where $\LF[\xpzc_{t},\ypzc_{t}]$ is defined over all $\{\xpzc_{t},\ypzc_{t}\} \in \setX\times  \setY$.
With these path-wise quantities\cite{Kobayashi:2015eca}, we obtain the population size of the organisms at time $t$,  the past phenotypic history of which is $\xpzc_{t}$ as 
\[
\popN_{t}^{\ypzc, \zpzc}[\xpzc_{t}] =e^{\LF[\xpzc_{t},\ypzc_{t}]}\ProbP_{F}[\xpzc_{t}||\zpzc_{t}]\popN_{0}^{\ypzc,\zpzc}.
\]
Because $\popN^{\ypzc,\zpzc}_{t}=\sum_{\xpzc_{t}}\popN_{t}^{\ypzc_{t}, \zpzc_{t}}[\xpzc_{t}]$, the cumulative fitness with sensing $\CMF^{\cs}[\ypzc_{t},\zpzc_{t}]$ is explicitly described as   
\begin{align}
\CMF^{\cs}[\ypzc_{t},\zpzc_{t}] &= \ln\frac{\popN_{t}^{\ypzc,\zpzc}}{\popN_{0}^{\ypzc,\zpzc}}=\ln\average{e^{\LF[\xpzc_{t},\ypzc_{t}]}}_{\ProbP_{F}[\xpzc_{t}||\zpzc_{t}]}.\label{eq:CMFF}
\end{align}
From this representation, the fitness can be described variationally \cite{Sughiyama:2015cf,Kobayashi:2015eca} as 
\begin{align}
\CMF^{\cs}[\ypzc_{t},\zpzc_{t}] &=\max_{\ProbP[\xpzc_{t}]}\left[\average{\LF[\xpzc_{t},\ypzc_{t}]}_{\ProbP[\xpzc_{t}]}-\KL{\ProbP[\xpzc_{t}]}{\ProbP_{F}[\xpzc_{t}||\zpzc_{t}]} \right],
\end{align}
where $\KLD[\ProbP||\ProbP']\defeq \sum_{\xpzc_{t}}\ProbP[\xpzc_{t}]\ln \ProbP[\xpzc_{t}]/\ProbP'[\xpzc_{t}]$ is the Kullback-Leibler divergence (relative entropy)\cite{Kullback:1951va,Cover:2012ub} between two path measures $\ProbP$ and $\ProbP'$.
\com{It should be noted that both KL divergence and Kramer's causal conditioning use the double bar $||$ but their meanings are different. }
The maximization is achieved with the time-backward retrospective path probability defined by 
\begin{align}
\ProbP^{\cs}_{B}[\xpzc_{t}|\ypzc_{t},\zpzc_{t}] &\defeq e^{\LF[\xpzc_{t},\ypzc_{t}]-\CMF^{\cs}[\ypzc_{t},\zpzc_{t}]}\ProbP_{F}[\xpzc_{t}||\zpzc_{t}] \com{=\frac{\popN_{t}^{\ypzc,\zpzc}[\xpzc_{t}]}{\popN_{t}^{\ypzc,\zpzc}} }. \label{eq:PBc}
\end{align}
If the phenotypic switching does not depend on the sensing signal as $\ProbP_{F}[\xpzc_{t}||\zpzc_{t}]=\ProbP_{F}[\xpzc_{t}]$, which corresponds to the bet-hedging by random phenotypic switching,  $\ProbP_{B}^{\cs}$is reduced to 
\begin{align}
\ProbP^{b}_{B}[\xpzc_{t}|\ypzc_{t}] &\defeq e^{\LF[\xpzc_{t},\ypzc_{t}]-\CMF^{b}[\ypzc_{t}]}\ProbP_{F}[\xpzc_{t}] = \frac{\popN_{t}^{\ypzc}[\xpzc_{t}]}{\popN_{t}^{\ypzc}},\label{eq:PBb}
\end{align}
where the superscript $b$ denotes bet-hedging and $\CMF^{b}[\ypzc_{t}] \defeq \ln\average{e^{\LF[\xpzc_{t},\ypzc_{t}]}}_{\ProbP_{F}[\xpzc_{t}]}$.
\com{Because $\popN_{t}^{\ypzc,\zpzc}[\xpzc_{t}]$ is the number of organisms with phenotypic history $\xpzc_{t}$ in the population at time $t$, the second equality in \eqnref{eq:PBc} indicates that $\ProbP^{s}_{B}[\xpzc_{t}|\ypzc_{t},\zpzc_{t}]$ is the fraction of the organisms that has phenotpyic history $\xpzc_{t}$.
This property of $\ProbP^{s}_{B}[\xpzc_{t}|\ypzc_{t},\zpzc_{t}]$ leads to an interpretation of $\ProbP^{s}_{B}[\xpzc_{t}|\ypzc_{t},\zpzc_{t}]$ as the probability of observing a certain phenotypic history $\xpzc_{t}$ under a realization of the environmental and signal histories $\ypzc_{t}$ and $\zpzc_{t}$ when we sample an organism in the population at time $t$ and track its phenotypic history in a time-backward manner, retrospectively\cite{Baake:2006ek,Wakamoto:2012hx,Sughiyama:2015cf,Lambert:2015bk,Kobayashi:2015eca}.}
Because the organisms grow more if their phenotypic histories are more adaptive than others for the given environmental realization $\ypzc_{t}$, the chance to observe a certain phenotypic history $\xpzc_{t}$ after selection under $\ypzc_{t}$ is biased to $\ProbP^{s}_{B}[\xpzc_{t}|\ypzc_{t},\zpzc_{t}]$ from the probability $\ProbP_{F}[\xpzc_{t}||\zpzc_{t}]$ with which the same phenotypic history is intrinsically generated.
Because the selected phenotypic histories strongly depend on the actual realization of the environmental history, $\ProbP^{s}_{B}[\xpzc_{t}|\ypzc_{t},\zpzc_{t}]$ is conditional on $\ypzc_{t}$.
\com{It should be noted that  $\ProbP_{B}^{\cs}[\xpzc_{t}|\ypzc_{t},\zpzc_{t}]$ is not necessarily causal, because of which we use the normal conditioning $|$\cite{Sughiyama:2015cf,Kobayashi:2015eca}.}

\subsection{Decomposition of fitness}
In order to understand the relation between  fitness and information obtained by sensing, 
we decompose the cumulative fitnesses into biologically relevant components. 
To obtain the decompositions, we first define a constant $\phi_{0}$ and a probability distribution $q_{0}(y)$ by using $\lf_{\max}(y)$ as $\phi_{0} \defeq - \ln \sum_{y}e^{-\lf_{\max}(y)}$ and $q_{0}(y)\defeq e^{\phi_{0}}e^{-\lf_{\max}(y)}$.
From these definitions, $\lf_{\max}(y)$ can be described as  $e^{\lf_{max}(y)}= e^{\phi_{0}}/q_{0}(y)$.
By defining $\ProbQ_{0}[\ypzc_{t}]\defeq \prod_{\tau=0}^{t-1}q_{0}(y_{\tau+1})$, $\ProbP_{\LF}[\ypzc_{t}||\xpzc_{t}]\defeq \prod_{\tau=0}^{t-1}\ProbT_{\LF}(y_{\tau+1}|x_{\tau+1})$, and $\Phi_{0}\defeq t \phi_{0}$, we obtain the following decomposition of $\LF$:
\begin{align}
\LF[\xpzc_{t},\ypzc_{t}] = \Phi_{0} + \ln \frac{\ProbP_{\LF}[\ypzc_{t}||\xpzc_{t}]}{\ProbQ_{0}[\ypzc_{t}]} \label{eq:DecombK},
\end{align}
where we use \eqnref{eq:ek}.
By defining $\LF_{\max}[\ypzc_{t}]\defeq \sum_{\tau=0}^{t-1}\lf_{max}(y_{\tau+1})=\Phi_{0}-\ln \ProbQ_{0}[\ypzc_{t}]$, $\LF[\xpzc_{t},\ypzc_{t}]$ can also be described as
\begin{align}
\LF[\xpzc_{t},\ypzc_{t}] = \LF_{\max}[\ypzc_{t}] + \ln \ProbP_{\LF}[\ypzc_{t}||\xpzc_{t}].
\end{align}

With eqs. (\ref{eq:PBb}) and (\ref{eq:DecombK}), for the bet-hedging problem, we obtain a decomposition of the fitness 
\begin{align}
\CMF^{b}[\ypzc_{t}] = \Phi_{0} - \ln \ProbQ_{0}[\ypzc_{t}] -\ln \frac{\ProbP^{b}_{B}[\xpzc_{t}| \ypzc_{t}]}{\ProbP_{\LF}[\ypzc_{t}||\xpzc_{t}]\ProbP_{F}[\xpzc_{t}]}.
\end{align}
For a given environmental statistics $\ProbQ[\ypzc_{t}]$, the fitness is represented by the ratio of the time-forward and time-backward path probabilities as 
\begin{align}
\CMF^{b}[\ypzc_{t}] = \CMF_{0}[\ypzc_{t}] -\ln \frac{\ProbP^{b}_{B}[\xpzc_{t}, \ypzc_{t}]}{\ProbP_{\LF}[\ypzc_{t}||\xpzc_{t}]\ProbP_{F}[\xpzc_{t}]},\label{eq:FDb_1}
\end{align}
where $\ProbP^{b}_{B}[\xpzc_{t},\ypzc_{t}]\defeq \ProbP^{b}_{B}[\xpzc_{t}|\ypzc_{t}]\ProbQ[\ypzc_{t}]$ is the time-backward joint probability of the phenotypic and environmental histories $\xpzc_{t}$ and $\ypzc_{t}$.
Here we also define 
\begin{align}
\CMF_{0}[\ypzc_{t}]\defeq \Phi_{0} + \ln \frac{\ProbQ[\ypzc_{t}]}{\ProbQ_{0}[\ypzc_{t}]} =\LF_{\max}[\ypzc_{t}]+\ln \ProbQ[\ypzc_{t}].\label{eq:Phi0}
\end{align}
If the organisms can perfectly foresee that the environmental state at time $\tau $ becomes $y_{\tau}$ and if they can choose the phenotype that allocates all metabolic resource to the environmental state $y_{\tau}$, the maximum replication rate at time $\tau$ reaches $e^{\lf_{\max}(y_{\tau})}$. 
Therefore, $\LF_{\max}[\ypzc_{t}]$ is interpreted as the maximum replication over an environmental path $\ypzc_{t}$ that can be achieved only when the organisms perfectly foresee what kind of environmental history will be realized in advance.
In contrast, $\ln \ProbQ[\ypzc_{t}]$ is the entropic loss of fitness due to the lack of knowledge of which environmental history will be realized\cite{Kussell:2005dg,Rivoire:2011fy}.
Therefore, $\CMF_{0}[\ypzc_{t}]$ is the maximum replication when the organisms cannot know which environmental state will be realized but know the statistics of the future environmental state.
The relevance of this interpretation and the biological meaning of some quantities such as $\Phi_{0}$ and $\ProbQ_{0}[\ypzc_{t}]$ are explicitly shown by using the FRs derived in the following sections.

For the case with the sensing signal, we can similarly obtain a decomposition of $\CMF^{\cs}[\ypzc_{t},\zpzc_{t}]$ as
\begin{align}
\CMF^{\cs}[\ypzc_{t},\zpzc_{t}] = \CMF_{0}[\ypzc_{t}]  -\ln \frac{\ProbP_{B}^{\cs}[\xpzc_{t},\ypzc_{t},\zpzc_{t}]}{\ProbP_{\LF}[\ypzc_{t}||\xpzc_{t}]\ProbP_{F}[\xpzc_{t}||\zpzc_{t}]\ProbQ[\zpzc_{t}|\ypzc_{t}]} \label{eq:FitnessDecombCommon}
\end{align}
where $\ProbP_{B}^{\cs}[\xpzc_{t},\ypzc_{t},\zpzc_{t}] \defeq \ProbP_{B}^{\cs}[\xpzc_{t}|\ypzc_{t},\zpzc_{t}]\ProbQ[\ypzc_{t},\zpzc_{t}]$ is the time-backward joint path probability among $\xpzc_{t}$, $\ypzc_{t}$, and $\zpzc_{t}$.
It should be noted that $\ProbP_{\LF}[\ypzc_{t}||\xpzc_{t}]\ProbP_{F}[\xpzc_{t}||\zpzc_{t}]\ProbQ[\zpzc_{t}|\ypzc_{t}]$ is not a joint path probability because of the circular noncausal dependency among $\xpzc_{t}$, $\ypzc_{t}$, and $\zpzc_{t}$.
Finally, by using the decomposition in \eqnref{eq:DecombK}, the time-backward conditional path probabilities, \eqnref{eq:PBb} and \eqnref{eq:PBc}, are reduced to
\begin{align}
\ProbP^{b}_{B}[\xpzc_{t}|\ypzc_{t}] &= \frac{\ProbP_{\LF}[\ypzc_{t}||\xpzc_{t}]]\ProbP_{F}[\xpzc_{t}]}{\ProbP_{\LF,F}[\ypzc_{t}]},\label{eq:PBb2}\\
\ProbP^{\cs}_{B}[\xpzc_{t}|\ypzc_{t},\zpzc_{t}] &= \frac{\ProbP_{\LF}[\ypzc_{t}||\xpzc_{t}]\ProbP_{F}[\xpzc_{t}||\zpzc_{t}]}{\ProbP_{\LF,F}[\ypzc_{t}|\zpzc_{t}]},\label{eq:PBc2}
\end{align}
where the normalization factors are 
\begin{align*}
\ProbP_{\LF,F}[\ypzc_{t}] &\defeq \sum_{\xpzc_{t}}\ProbP_{\LF}[\ypzc_{t}||\xpzc_{t}]]\ProbP_{F}[\xpzc_{t}],\\ 
\ProbP_{\LF,F}[\ypzc_{t}|\zpzc_{t}] &\defeq \sum_{\xpzc_{t}}\ProbP_{\LF}[\ypzc_{t}||\xpzc_{t}]\ProbP_{F}[\xpzc_{t}||\zpzc_{t}].
\end{align*}
Because $\ProbP_{F}[\xpzc_{t}]$ is the probability to intrinsically generate the repertoire of phenotypic histories in the population and $\ProbP_{\LF}[\ypzc_{t}||\xpzc_{t}]$ is the probability that an organism with a phenotypic history $\xpzc_{t}$ allocates resources to each history of environment $\ypzc_{t}$,  the normalization factor $\ProbP_{\LF,F}[\ypzc_{t}]$ can be interpreted as the marginal resource allocation to the environmental history $\ypzc_{t}$ at the population level.
$\ProbP_{\LF,F}[\ypzc_{t}|\zpzc_{t}]$ can similarly be interpreted as the population-level resource allocation to $\ypzc_{t}$ when signal history $\zpzc_{t}$ is received.
By using FRs in the next section, we clarify that $\ProbP_{F}[\xpzc_{t}]$ and $\ProbP_{\LF,F}[\ypzc_{t}|\zpzc_{t}]$ also have meaning as the conjugate environment under which the given strategy $\{\ProbT_{F}, \ProbT_{\LF}\}$ becomes optimal.

\section{Causal FRs for bet-hedging strategy}\label{sec:bethedging}
By rearranging the decomposition of $\CMF^{b}[\ypzc_{t}]$ in \eqnref{eq:FDb_1}, 
we can immediately obtain a detailed causal FR for fitness difference $\CMF_{0}[\ypzc_{t}]-\CMF^{b}[\ypzc_{t}]$ as
\begin{align}
e^{-(\CMF_{0}[\ypzc_{t}]-\CMF^{b}[\ypzc_{t}])} =\frac{\ProbP_{\LF}[\ypzc_{t}||\xpzc_{t}]\ProbP_{F}[\xpzc_{t}]}{\ProbP_{B}^{b}[\xpzc_{t},\ypzc_{t}]}= \frac{\ProbP_{\LF,F}[\ypzc_{t}]}{\ProbQ[\ypzc_{t}]}, \label{eq:DFRb}
\end{align}
where we use \eqnref{eq:PBb2} to obtain the last equality.
The first equality means that the fitness difference is the log ratio of time-forward and time-backward path probabilities $\ProbP_{\LF}[\ypzc_{t}||\xpzc_{t}]\ProbP_{F}[\xpzc_{t}]$ and $\ProbP_{B}^{b}[\xpzc_{t},\ypzc_{t}]$.
$\ProbP_{\LF}[\ypzc_{t}||\xpzc_{t}]\ProbP_{F}[\xpzc_{t}]$ is the time-forward probability of observing an organism that  takes phenotypic history $\xpzc_{t}$ and then allocates $\ProbP_{\LF}[\ypzc_{t}||\xpzc_{t}]$ of metabolic resources to $\ypzc_{t}$ \textit{a priori} to selection by conducting time-forward tracking of the histories.
$\ProbP_{B}^{b}[\xpzc_{t},\ypzc_{t}]$ is the time-backward probability of observing the realization of environmental history $\ypzc_{t}$ and the time-backward phenotypic history $\xpzc_{t}$ \textit{a posteriori} to selection by conducting time-backward tracking of the histories.
The second equality also indicates that the fitness difference is the log ratio of the percentage of resource allocated to environmental history $\ypzc_{t}$ at the population level and the probability of observing environmental history $\ypzc_{t}$.

By averaging \eqnref{eq:DFRb} with respect to $\ProbP_{B}^{b}[\xpzc_{t},\ypzc_{t}]$ or $\ProbQ[\ypzc_{t}]$, we can derive an integral FR as 
\begin{align}
\average{e^{-(\CMF_{0}[\ypzc_{t}]-\CMF^{b}[\ypzc_{t}])}}_{\ProbQ[\ypzc_{t}]}= 1\label{eq:IFRb}.
\end{align}
If we average \eqnref{eq:DFRb} after taking the logarithm of both sides, we obtain an average FR as  
\begin{align}
\average{\CMF^{b}}_{\ProbQ} = \average{\CMF_{0}}_{\ProbQ} -\KLD^{b}_{\loss},\label{eq:AFRb}
\end{align}
where $\average{\CMF^{b}}_{\ProbQ} \defeq \average{\CMF^{b}[\ypzc_{t}]}_{\ProbQ[\ypzc_{t}]}$, $\average{\CMF_{0}}_{\ProbQ} \defeq \average{\CMF_{0}[\ypzc_{t}]}_{\ProbQ[\ypzc_{t}]}$, and 
\begin{align}
\KLD^{b}_{\loss} &\defeq \KL{\ProbP^{b}_{B}[\xpzc_{t}, \ypzc_{t}]}{\ProbP_{\LF}[\ypzc_{t}||\xpzc_{t}]\ProbP_{F}[\xpzc_{t}]} \\
&=\KL{\ProbQ[\ypzc_{t}]}{\ProbP_{\LF,F}[\ypzc_{t}]} \label{eq:Dloss}
\end{align}
Because of the non-negativity of the relative entropy $\KLD^{b}_{\loss}$, we can easily see that $ \average{\CMF_{0}}_{\ProbQ}$ is an upper bound of the average fitness $\average{\CMF^{b}}_{\ProbQ}$ of a bet-hedging strategy: 
\begin{align}
\average{\CMF_{0}}_{\ProbQ} \ge \max_{\{\ProbT_{F},\ProbT_{\LF}\}}\average{\CMF^{b}}_{\ProbQ}  ,\label{eq:AFRbBound}
\end{align}
where we use the fact that $\CMF_{0}[\ypzc_{t}]$ is dependent neither on the phenotype switching strategy $\ProbT_{F}$ nor on the metabolic allocation strategy $\ProbT_{\LF}$.
\com{These FRs are basically the same as those we derived in our previous work by using a variational approach. 
It should be also noted that the FRs derived by Mustonen and L\"assig\cite{Mustonen:2010ig} are different from ours because their relations are based on a model describing the dynamics of an ensemble of populations whereas ours is one describing the dynamics of a population.}

\subsection{Biological meaning of $\CMF_{0}$ and $\Phi_{0}$}
From \eqnref{eq:Phi0}, the upper bound of the average fitness, $\average{\CMF_{0}}_{\ProbQ}$ admits two different representations:
\begin{align}
\average{\CMF_{0}}_{\ProbQ} = \average{\LF_{\max}}_{\ProbQ}-\mathcal{S}[\ProbQ]=\Phi_{0} +  \KL{\ProbQ}{\ProbQ_{0}},
\end{align}
where $\mathcal{S}[\ProbQ] \defeq - \average{\ln \ProbQ[\ypzc_{t}]}_{\ProbQ[\ypzc_{t}]}$ is the entropy of $\ProbQ[\ypzc_{t}]$ and $\average{\LF_{\max}}_{\ProbQ}$ is the average fitness under environmental statistics $\ProbQ[\ypzc_{t}]$ that is attained only when organisms have perfect knowledge of the future environment.
Therefore, the first equality means that the randomness of the environment quantified by the entropy $\mathcal{S}[\ProbQ]$ works as the inevitable loss of fitness due to the lack of knowledge of which environmental history will be realized in the future.
If the environment fluctuates more unpredictably, we have a higher $\mathcal{S}[\ProbQ]$ and a lower upper bound of the average fitness.
Note that these properties of $\average{\CMF_{0}}_{\ProbQ}$ have been pointed out previously and repeatedly \cite{Haccou:1995tf,Kussell:2005dg,Rivoire:2011fy}.

\begin{figure}
\includegraphics[width=\linewidth]{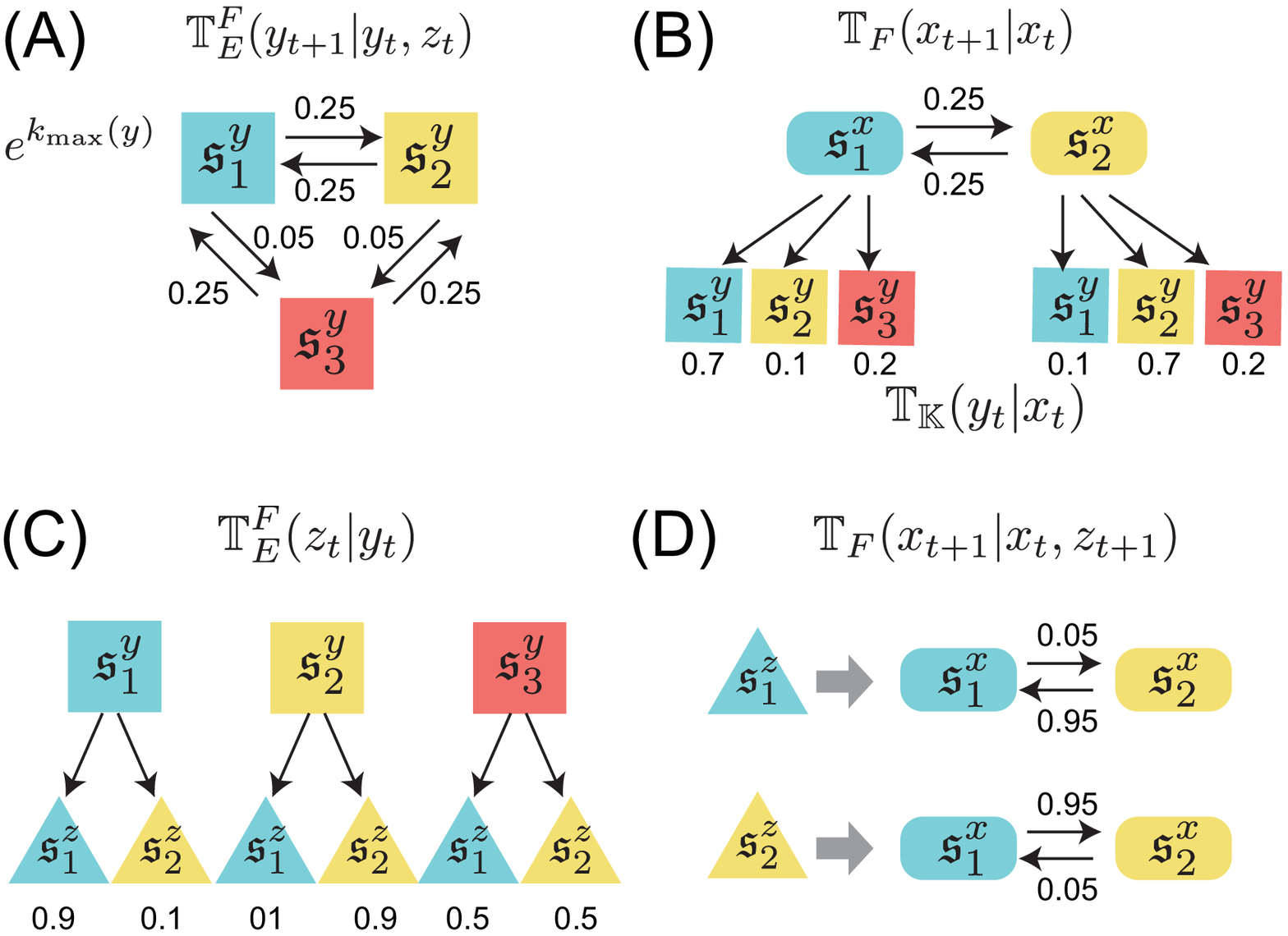}
\caption{\label{fig2} 
Schematic representations of environmental dynamics, phenotypic switching, and sensing used for numerical verification of FRs in Figs. 3 and 5.
(A) Characteristics of the environment. The environment has three states: $y_{t} \in \sety=\{\ely_{1}, \ely_{2}, \ely_{3}\}$. 
$\ely_{1}$ and $\ely_{2}$ are nutrient rich states that have different sources of nutrients. 
$\ely_{3}$ is a nutrient-poor state under which replication of the organisms is severely restricted.
The maximum growth under each environment is $e^{\lf_{\max}(\ely_{1})}=3.2$, $e^{\lf_{\max}(\ely_{2})}=3.2$, and $e^{\lf_{\max}(\ely_{3})}=0.4$.
The transition rates $\ProbT_{E}^{F}(y_{t+1}|y_{t})$ between the states are shown on the arrows.
The environmental states usually fluctuate between $\ely_{1}$ and $\ely_{2}$ but occasionally flip to $\ely_{3}$ with a $5\%$ chance.
(B) Characteristics of phenotypic states without sensing. 
Organisms have two phenotypic states: $x_{t} \in \setx = \{\elx_{1}, \elx_{2}\}$. 
The transition rates $\ProbT_{F}(x_{t+1}|x_{t})$ between the states are shown on the arrows. 
$\elx_{1}$ is the phenotypic state allocating more resources to the environmental state $\ely_{1}$ ($70 \%$) than $\elx_{2}$ ($10 \%$) whereas $\elx_{2}$ allocates more to $\elx_{2}$ ($70 \%$) than $\elx_{1}$ ($10 \%$).
 Both states allocate $20 \%$ of the resources to the starving environmental state $\ely_{3}$.
(C) Characteristic of sensing signal. The sensing signal has two states as $z_{t} \in \setz =\{\elz_{1}, \elz_{2}\}$.
When the environmental state is $\ely_{1}$ ($\ely_{2}$), the organisms obtain $\elz_{1}$ ($\elz_{2}$) as the sensing signal with a $90\%$ accuracy. 
If the environment is in $\ely_{3}$, the sensing signal produces $\elz_{1}$ or $\elz_{2}$ with equal probability.
(D) Characteristic of phenotypic switching with sensing. 
When the organism obtains the sensing signal $\elz_{1}$ ($\elz_{2}$), it switches its phenotypic state into $\elx_{1}$ ($\elx_{2}$) $95 \%$ of the time.
}
\end{figure}

The meaning of $\Phi_{0}$ and $\ProbQ_{0}[\ypzc_{t}]$ in the second equality becomes explicit by considering the minimization of $\average{\CMF_{0}}_{\ProbQ} $ with respect to $\ProbQ[\ypzc_{t}]$ as follows:
\begin{align}
\Phi_{0} =\min_{\ProbQ}\average{\CMF_{0}[\ypzc_{t}]}_{\ProbQ[\ypzc_{t}]} \ge \min_{\ProbQ}\max_{\{\ProbT_{F},\ProbT_{\LF}\}}\average{\CMF^{b}[\ypzc_{t}]}_{\ProbQ[\ypzc_{t}]} ,
\end{align}
where we use 
\begin{align}
 \min_{\ProbQ}\average{\CMF_{0}[\ypzc_{t}]}_{\ProbQ[\ypzc_{t}]}=\Phi_{0} +  \min_{\ProbQ}\KLD[\ProbQ||\ProbQ_{0}] =\Phi_{0}.
 \end{align}
This relation indicates that $\Phi_{0}$ is the minimum of the maximum average fitness and that $\ProbQ_{0}[\ypzc_{t}]$ is the worst environment for the organisms under which the maximum average fitness is minimized. 
This min-max characterization of $\Phi_{0}$ and $\ProbQ_{0}[\ypzc_{t}]$ has been clarified in the context of game theory with a matrix formulation\cite{Pugatch:2013va}.

\subsection{Verification of FRs for fitness}\label{ssec:VerFRbf}
Equations (\ref{eq:DFRb}--\ref{eq:AFRb}) indicate that, under quite general situations, the fitness difference $\CMF_{0}[\ypzc_{t}]-\CMF^{b}[\ypzc_{t}]$  satisfies the FRs \cite{Kobayashi:2015eca} as the entropy production does in stochastic thermodynamics\cite{Seifert:2012es,Sagawa:2012wi}.
To demonstrate the relations, we consider an organism with two phenotypic states growing in a Markovian environment with three states as depicted in Fig. 2 (A) and (B).
The three environmental states, $\ely_{1}$, $\ely_{2}$, and $\ely_{3}$, describe nutrient A rich, nutrient B rich, and nutrient-poor conditions, respectively (Fig 2(A)).
The two phenotypic states, $\elx_{1}$ and $\elx_{2}$, employ strategies to allocate $70\%$ of the metabolic resources to $\ely_{1}$ and $\ely_{2}$, respectively.
Both states allocate $10\%$ resources to the rest of two states (see Appendix \ref{ap:num} for more details).
\com{This setting abstractly and simply represents the fact that organisms generally have much less phenotypic and sensing states than possible environmental states because of the limited physical complexity of the organisms\cite{comment1}.}

Figures 3 (A) and (B) show the population dynamics of the organisms with the phenotypic states $\elx_{1}$ and $\elx_{2}$ under two different realizations of the environmental history alongside $\CMF^{b}[\ypzc_{t}]$ and $\CMF_{0}[\ypzc_{t}]$.
Depending on the actual realization of the environment, the relative relations among $\popN_{t}^{\ypzc}(\elx_{1})$, $\popN_{t}^{\ypzc}(\elx_{2})$, $e^{\CMF^{b}[\ypzc_{t}]}=\popN_{t}^{\ypzc}$, and $e^{\CMF_{0}[\ypzc_{t}]}$ change over time stochastically.
In Fig 3 (A), $e^{\CMF_{0}[\ypzc_{t}]}$ is mostly greater than $\popN_{t}^{\ypzc}=e^{\CMF^{b}_{t}[\ypzc_{t}]}$, which reflects the average relation $\average{\CMF_{0}}_{\ProbQ} \ge \average{\CMF^{b}_{t}}_{\ProbQ}$.
On the contrary, $\popN_{t}^{\ypzc}$ frequently becomes greater than $e^{\CMF_{0}[\ypzc_{t}]}$ in Fig 3 (B).
Figures 3 (C) and (D) show that the environmental fluctuation induces a large fluctuation in both $\CMF^{b}[\ypzc_{t}]$ and $\CMF_{0}[\ypzc_{t}]$. 
As shown in Fig 3 (E), although most environmental fluctuations result in positive fitness differences, that is, $\CMF_{0}[\ypzc_{t}]-\CMF^{b}[\ypzc_{t}]>0$, rare environmental fluctuations lead to a negative fitness difference in a finite time interval, meaning that the fitness of the suboptimal strategy $\CMF^{b}[\ypzc_{t}]$ outperforms the average upper bound $\CMF_{0}[\ypzc_{t}]$.
This is analogue to the reversed heat flow in a small thermal system\cite{Jarzynski:2004ia}.
Such rare events are balanced to satisfy the integral FR in \eqnref{eq:IFRb} as verified numerically in Fig 3 (F).

\begin{figure}[H]
\includegraphics[width=\linewidth]{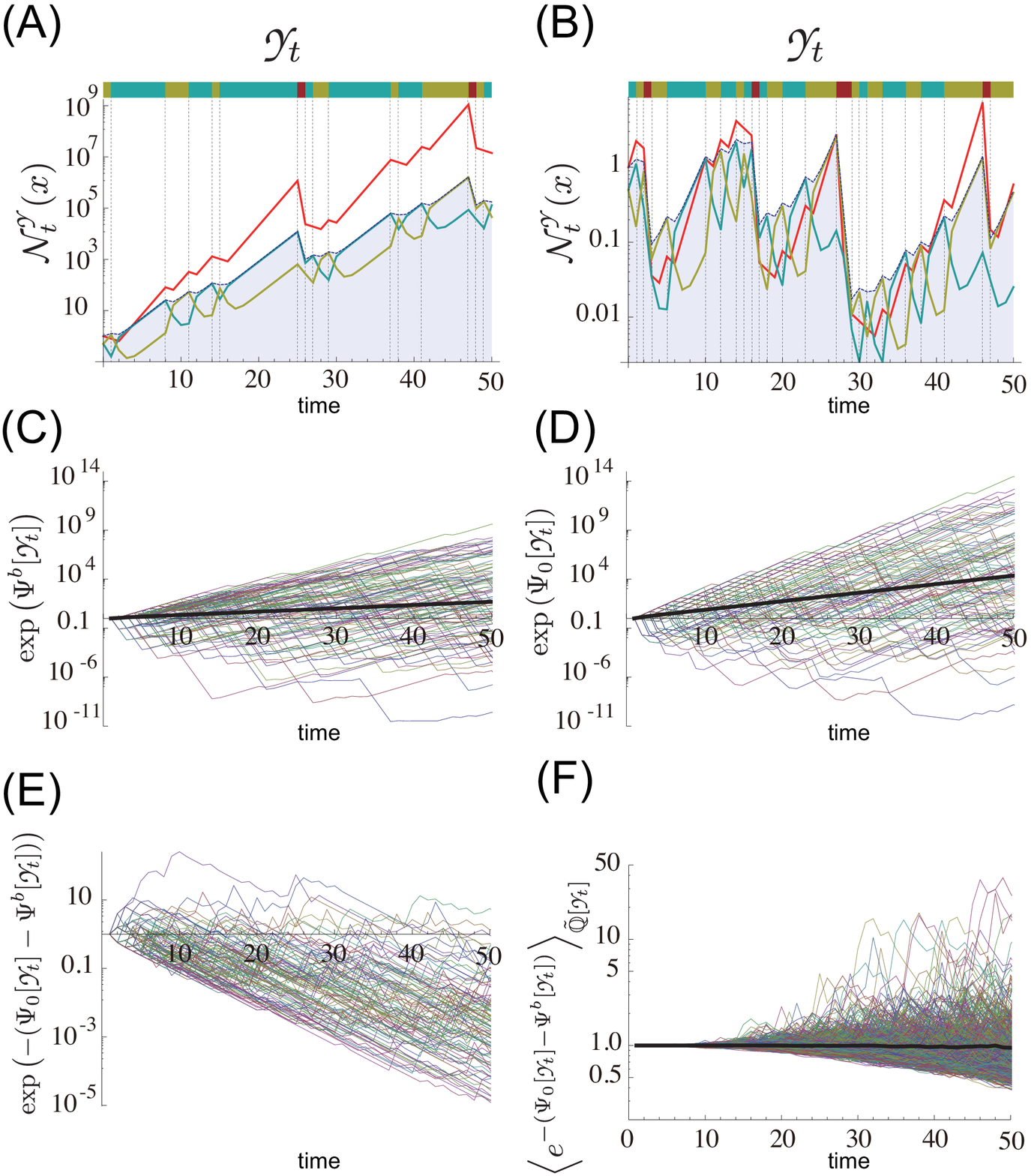}
\caption{\label{fig3} 
Numerical simulation of the population dynamics without sensing defined in Fig. 2.
(A, B)  Two sample histories of environment $\ypzc_{t}$ and the dynamics of $\popN_{t}^{\ypzc}(\elx_{1})$, $\popN_{t}^{\ypzc}(\elx_{2})$, $\popN_{t}^{\ypzc}=e^{\CMF^{b}[\ypzc_{t}]}$, and $e^{\CMF_{0}[\ypzc_{t}]}$ under $\ypzc_{t}$  \com{obtained by solving \eqnref{eq:popdyn}}. 
\com{We set $\popN_{0}^{\ypzc} =1$.}
The color bars on the graphs represent the state of the environment at each time point. 
The correspondence between the colors and environmental states is as shown in Fig. 2 (A). 
Cyan and yellow lines represent $\popN_{t}^{\ypzc}(\elx_{1})$ and $\popN_{t}^{\ypzc}(\elx_{2})$, respectively.
The dashed blue line with filled grey style is  $\popN_{t}^{\ypzc}$. 
The red line is $\CMF_{0}[\ypzc]$.
(C) Stochastic behaviors of population fitness $\CMF^{b}[\ypzc_{t}]$ for $100$ independent samples of the environmental histories. 
Each colored thin line represents $e^{\CMF^{b}[\ypzc_{t}]}=\popN_{t}^{\ypzc}/\popN_{0}^{\ypzc}$ for each realization of environmental history. 
The thick black line is $e^{\average{\CMF_{b}[\ypzc_{t}]}_{\ProbQ[\ypzc_{t}]}}$.
(D) Stochastic behaviors of $\CMF_{0}[\ypzc_{t}]$ for the same $100$ samples of environmental histories as those in (C). 
Each colored thin line represents $e^{\CMF_{0}[\ypzc_{t}]}$ for each realization of the same environmental history as in (C). 
(E) Stochastic behaviors of $e^{-(\CMF_{0}[\ypzc_{t}]-\CMF^{b}[\ypzc_{t}])}$ for the same $100$ samples of environmental histories as those in (C). 
Each colored thin line represents $e^{-(\CMF_{0}[\ypzc_{t}]-\CMF^{b}[\ypzc_{t}])}$ for each realization of the same environmental history as in (C).
(F) $\average{e^{-(\CMF_{0}[\ypzc_{t}]-\CMF^{b}[\ypzc_{t}])}}_{\ProbQ[\ypzc_{t}]}$ calculated empirically by the numerical simulations of sample paths of $e^{-(\CMF_{0}[\ypzc_{t}]-\CMF^{b}[\ypzc_{t}])}$.
\com{The thick line is the average of $10^{8}$ independent samples of environmental histories.
To illustrate the fluctuation, the $10^{8}$ histories are dissected into 100 groups of histories, each contains $10^{5}$ histories.
Each thin line is obtained as the average of $10^{5}$ histories in each group.}
}
\end{figure}

\section{Biological meaning of the Fitness FRs}\label{sec:BMFRbh}
In the original detailed FR over paths\cite{Sagawa:2012wi}, the entropy production is the log ratio of the path probability of a system's trajectory and its time reversal.
The average entropy production attains its minimum $0$ only when the time reversibility of the system holds in the sense that the probabilities of observing the time-forward and the time-reversed trajectories are equal.
Thus, the FRs are related to the extent of the time reversibility of the system.
In contrast, the detailed FRs for the fitness difference (\eqnref{eq:DFRb}) is the log ratio between the probability $\ProbQ[\ypzc_{t}]$ of observing the environmental history and the percentage $\ProbP_{\LF,F}[\ypzc_{t}]$ of the marginal resource allocation to $\ypzc_{t}$, or that between the time-backward  path probability $\ProbP_{B}^{b}[\xpzc_{t},\ypzc_{t}]$ and time-forward path probability $\ProbP_{\LF}[\ypzc_{t}||\xpzc_{t}]\ProbP_{F}[\xpzc_{t}]$.
By investigating the FRs, we clarify a dualistic structure, a conjugacy of these quantities, and the time-reversal condition for the equality attained in \eqnref{eq:AFRbBound}.


\subsection{Dualistic relation between strategy and environment}
The average FR in \eqnref{eq:AFRb} implies that the maximization of $\average{\CMF^{b}}_{\ProbQ}$ with respect to the strategies is dual to the minimization of the relative entropy $\KLD_{\loss}^{b}$ as follows: 
\begin{align}
\{\ProbT_{F}^{\dagger},\ProbT_{\LF}^{\dagger}\} &\defeq\arg\max_{\{\ProbT_{F},\ProbT_{\LF}\}}\average{\CMF^{b}}_{\ProbQ}=\arg \min_{\{\ProbT_{F},\ProbT_{\LF}\}}\KLD_{\loss}^{b},\label{eq:optb}
\end{align}
because $\CMF_{0}[\ypzc_{t}]$ is independent of $\ProbP_{F}$ and $\ProbP_{\LF}$.
This duality indicates that maximizing the average fitness by choosing the best strategy is equivalent to the organisms to implicitly learn and prepare for the environmental statistics $\ProbQ[\ypzc_{t}]$ so that the marginal resource allocation $\ProbP_{\LF, F}[\ypzc_{t}]$ to $\ypzc_{t}$ becomes close to $\ProbQ[\ypzc_{t}]$ because $\KLD_{\loss}^{b}=\KLD[\ProbQ[\ypzc_{t}]||\ProbP_{\LF, F}[\ypzc_{t}]]$.
The upper bound of the average fitness in \eqnref{eq:AFRbBound} is achieved as $\CMF^{b}[\ypzc_{t}]=\CMF_{0}[\ypzc_{t}]$ if and only if $\{\ProbT_{F}^{\dagger},\ProbT_{\LF}^{\dagger}\}$ satisfies $\ProbQ[\ypzc_{t}]=\ProbP_{\LF, F}^{\dagger}[\ypzc_{t}]$ where $\ProbP_{\LF, F}^{\dagger}[\ypzc_{t}] \defeq \sum_{\xpzc_{t}}\ProbT_{\LF}^{\dagger}[\ypzc_{t}||\xpzc_{t}]\ProbT_{F}^{\dagger}[\xpzc_{t}]$, meaning that the environmental statistics and the marginal resource allocation match perfectly.

\subsection{Meaning of $\ProbP_{F,\LF}[\ypzc_{t}]$ as conjugate environment}
For a given environment, the strategy that achieves the bound may not always exist.
In contrast, for a given pair of strategies $\{\ProbT_{F},\ProbT_{\LF}\}$, there always exists the environment $\ProbQ^{\dagger}[\ypzc_{t}]$ under which the pair achieves the bound
\begin{align}
\average{\CMF_{0}}_{\ProbQ^{\dagger}} = \average{\CMF^{b}[\ypzc_{t}]}_{\ProbQ^{\dagger}[\ypzc_{t}]} \ge \max_{\{\ProbT'_{F},\ProbT'_{\LF}\}}\average{{\CMF^{b}}'[\ypzc_{t}]}_{\ProbQ^{\dagger}[\ypzc_{t}]}, \label{eq:bhbound}
\end{align}
and 
\begin{align}
\{\ProbT_{F},\ProbT_{\LF}\} = \arg\max_{\{\ProbT'_{F},\ProbT'_{\LF}\}}\average{{\CMF^{b}}'[\ypzc_{t}]}_{\ProbQ^{\dagger}[\ypzc_{t}]},
\end{align}
is satisfied.
Because of the duality shown in \eqnref{eq:optb}, $\ProbQ^{\dagger}[\ypzc_{t}]$  is explicitly obtained as $\ProbQ^{\dagger}[\ypzc_{t}]=\ProbP_{F,\LF}[\ypzc_{t}]$.
Therefore, $\ProbP_{\LF, F}[\ypzc_{t}]$ can also be regarded as the conjugate environment to the strategy $\{\ProbT_{F},\ProbT_{\LF}\}$ or $\{\ProbP_{F},\ProbP_{\LF}\}$ under which they are optimal.  
Therefore, the fitness difference in \eqnref{eq:DFRb} is the log ratio of the actual environmental statistics $\ProbQ[\ypzc_{t}]$ and that of the conjugate environment $\ProbQ^{\dagger}[\ypzc_{t}]=\ProbP_{\LF, F}[\ypzc_{t}]$ of the given strategy.

$\CMF^{b}[\ypzc_{t}]$ is bounded by $\CMF_{0}[\ypzc_{t}]$ on average, and therefore, the optimal strategy that attains $\CMF_{0}[\ypzc_{t}]$ cannot be invaded by any other strategy if we consider an infinitely large population and the asymptotic dynamics. 
Thus, the optimal strategy is a version of ESS in a fluctuating environment.
Within a finite time interval, however, $\CMF^{b}[\ypzc_{t}]$ becomes greater than  $\CMF_{0}[\ypzc_{t}]$ under certain realizations of environment $\ypzc_{t}$ that satisfy $\ProbQ[\ypzc_{t}]<\ProbQ^{\dagger}[\ypzc_{t}]$.
In stochastic thermodynamics, such realizations correspond to the temporal reversed heat flow in a small thermal system\cite{Jarzynski:2004ia}.
In biology, they temporally violate ESS because a suboptimal strategy $\{\ProbT_{F},\ProbP_{\LF}\}$ under $\ProbQ$ can outperform the optimal strategy or near optimal strategies with the aid of the environmental fluctuation within a finite time interval. 
The integral FR in \eqnref{eq:IFRb} tells us that the violation of ESS can always occur with a small but finite probability in a finite time interval. 
If the population size of the optimal strategy is finite, such violation can leads to extinction of the optimal population with a finite probability\cite{King:2007fi}.
Moreover, the detailed FR in \eqnref{eq:DFRb} implies that a greater violation can occur under a realization of environmental history $\ypzc_{t}$  if $\ProbQ[\ypzc_{t}]$ is small but $\ProbQ^{\dagger}[\ypzc_{t}]$ is large.
This fact can be intuitively understood as follows: the greater violation is induced by the environmental history $\ypzc_{t}$ that rarely occurs in the actual environmental statistics $\ProbQ[\ypzc_{t}]$ but is adaptive and advantageous for the given strategy.
A crucial fact is that this intuitive understanding is supported by a quantitative relation as in \eqnref{eq:DFRb}.
Furthermore, the detailed FR suggests that greater violation can occur for specific suboptimal strategies than for others if $\ProbQ[\ypzc_{t}]$ contains many rare environmental histories.
In contrast, if $\ProbQ[\ypzc_{t}]$ is perfectly random as $\ProbQ[\ypzc_{t}]=\const$, the extent of the violation is limited and all the suboptimal strategies have only an even chance of violation.
This implies that structured environmental fluctuation promotes impactful violation by some specific suboptimal strategies.

\subsection{Time reversibility and optimality}
If $\CMF^{b}[\ypzc_{t}]=\CMF_{0}[\ypzc_{t}]$ holds, moreover, the time-backward probability $\ProbP_{B}^{b,\dagger}[\xpzc_{t},\ypzc_{t}]$ and the time-forward path probability $\ProbP_{\LF}^{\dagger}[\ypzc_{t}||\xpzc_{t}]\ProbP_{F}^{\dagger}[\xpzc_{t}]$ become equal:
\begin{align}
\ProbP_{B}^{b,\dagger}[\xpzc_{t},\ypzc_{t}]=\ProbP_{B}^{b,\dagger}[\xpzc_{t}|\ypzc_{t}]\ProbQ[\ypzc_{t}]=\ProbP_{\LF}^{\dagger}[\ypzc_{t}||\xpzc_{t}]\ProbP_{F}^{\dagger}[\xpzc_{t}]
\end{align}
where we use the first equality in \eqnref{eq:Dloss}. 
Marginalization of this equality with respect to $\ypzc_{t}$ indicates that the marginalized time-backward path probability $\ProbP_{B}^{b}[\xpzc_{t}]$ satisfies the consistency condition $\ProbP_{B}^{b,\dagger}[\xpzc_{t}]=\ProbP_{F}^{\dagger}[\xpzc_{t}]$  as shown previously\cite{Kobayashi:2015eca}.
Thus, the optimal strategies to achieve the bound have time reversibility in the sense that the ensembles of the time-forward and time-backward phenotypic histories are indistinguishable without knowing the actual environmental history that the population experienced.
While the definition of time reversibility is different, this result is closely related to the fact that the average entropy production attains $0$ when the time reversibility is satisfied.
Biologically, this result is quite important because we can evaluate the optimality of an organism in a changing environment by just observing its phenotypic dynamics without directly measuring the environment that the organisms experience.
It should be noted, however, that $\ProbP_{B}^{b,\dagger}[\xpzc_{t}]=\ProbP_{F}^{\dagger}[\xpzc_{t}]$ is a necessary but not sufficient condition for $\CMF^{b}[\ypzc_{t}]=\CMF_{0}[\ypzc_{t}]$.

\subsection{Achievability of the fitness upper bound}
The duality \eqnref{eq:optb} indicates that the achievability of the bound in \eqnref{eq:AFRbBound} depends on the actual property of the environmental statistics and biological constraints on the selectable strategies.
If the environment is a time-homogeneous Markov process as $\ProbQ[\ypzc_{t}]=\prod_{\tau=0}^{t-1}\ProbT_{E}(y_{\tau+1}|y_{\tau})q(y_{0})$ and if the number of possible phenotypic states is the same as that of the environmental states, that is, $\# \setX=\# \setY$, 
the bound can be achieved by a pair of strategies: 
\begin{align}
\{\ProbT_{F}^{\dagger}(x'|x),\ProbT_{\LF}^{\dagger}(y|x)\} =\{\ProbT_{E}(x'|x), \delta_{x,y}\}
\end{align}
where $\ProbT_{E}(x'|x)\defeq\left.\ProbT_{E}(y_{\tau+1}|y_{\tau})\right|_{y_{\tau+1}=x', y_{\tau}=x}$ and $\delta_{x,y}$ is the Kronecker delta.
This pair is equivalent to the optimal bet-hedging strategy in Kelly's horse race gambling because $\ProbT_{\LF}^{\dagger}(y|x)=\delta_{x,y}$ means that organisms can survive only when their phenotypic state matches the current environmental state; they die out otherwise.
To be specific, we call $\ProbT_{\LF}(y|x)=\delta_{x,y}$ Kelly's strategy of metabolic allocation.
Under biologically realistic constraints, however, Kelly's strategy cannot be the optimal strategy because the possible phenotypic states are usually much fewer than those of the environment as in Fig 2 and 3.
Allocating all resources to a specific environment easily leads to extinction if the phenotypic states cannot cover all the possible environmental states.
With a limited capacity in possible phenotypic states, $\KLD_{\loss}^{b}=0$ can be attained only if the environmental fluctuation has a hidden structure the dimensionality of which is sufficiently low.
This is manifested by the fact that the conjugate environment $\ProbQ^{\dagger}$ is the environment with a hidden dynamics $\ProbP_{F}[\xpzc'_{t}]$ that generates the actual environmental history as $\ProbQ^{\dagger}[\ypzc_{t}]=\sum_{\xpzc'_{t}}\ProbP_{\LF}[\ypzc_{t}|\xpzc'_{t}]\ProbP_{F}[\xpzc'_{t}]$.
Because such a low dimensional structure may not always exist, however,
\begin{align*}
\KLD_{\loss}^{b^{\dagger}} \defeq  \min_{\{\ProbT_{F},\ProbT_{\LF}\}}\KLD_{\loss}^{b}
\end{align*}
is generally not zero but finite and positive under a biological constraint that the possible phenotypic states are fewer than  environmental ones.  
Therefore, $\average{\CMF_{0}}_{\ProbQ}$ is generally attained only by a Darwinian demon that cannot perfectly foresee the future environment but has sufficient capacity in its phenotypic properties to perfectly learn and prepare for the environmental fluctuation $\ProbQ[\ypzc_{t}]$.

Even when the bound is not achieved, $\KLD_{\loss}^{b^{\dagger}}$ has explicit meaning as the component in the environmental fluctuation that cannot be learned or represented by the dynamics of the cell's strategy.
For example, when $\ProbT_{F}$ is memoryless and $\ProbQ[\ypzc_{t}]$ is a stationary Markov process with the stationary probability $q(y)$, that is, $\ProbQ[\ypzc_{t}]=\prod_{\tau=0}^{t-1}\ProbT_{E}^{F}(y_{\tau+1}|y_{\tau})q(y_{0})$, then 
\begin{align*}
\KLD_{\loss}^{b^{\dagger}} &= \min_{\{\ProbT_{F},\ProbT_{\LF}\}} \KLD_{\loss}^{b}\\
&=\min_{\{\ProbT_{F},\ProbT_{\LF}\}} \average{\ln \frac{\ProbT_{E}^{F}(y'|y)q(y)}{\average{\ProbT_{\LF}(y'|x)}_{\ProbT_{F}(x)}}}_{\ProbT_{E}^{F}(y'|y)q(y)}\\
&=\min_{\{\ProbT_{F},\ProbT_{\LF}\}}\left[ \MI^{x_{t}; x_{t+1}}+\KL{q(y)}{\average{\ProbT_{\LF}(y|x)}_{\ProbT_{F}(x)}} \right]\\
&=\MI^{x_{t}; x_{t+1}},
\end{align*}
where $\MI^{x_{t}; x_{t+1}} \defeq \KL{\ProbT_{E}^{F}(y'|y)q(y))}{q(y)q(y')}$ is the mutual information that measures the correlation of environmental states between consecutive time points, which cannot be learned or mimicd by the memoryless phenotypic switching.

\section{FRs with signal sensing}\label{sec:FRs}
Next, we consider the case in which the organisms can exploit the information obtained from the sensing signal.
The fitness decomposition in \eqnref{eq:FitnessDecombCommon} can be rearranged as
\begin{align}
\CMF^{\cs}[\ypzc_{t},\zpzc_{t}] &= \CMF_{0}[\ypzc_{t}] + i[\ypzc_{t}; \zpzc_{t}] -\ln \frac{\ProbP_{B}^{\cs}[\xpzc_{t},\ypzc_{t},\zpzc_{t}]}{\ProbP_{\LF}[\ypzc_{t}||\xpzc_{t}]\ProbP_{F}[\xpzc_{t}||\zpzc_{t}]\ProbQ[\zpzc_{t}]} \\
 &= \CMF_{0}[\ypzc_{t}] + i[\ypzc_{t}; \zpzc_{t}] -\ln \frac{\ProbQ[\ypzc_{t},\zpzc_{t}]}{\ProbP_{\LF,F}[\ypzc_{t}|\zpzc_{t}]\ProbQ[\zpzc_{t}]},
\end{align}
where  $i[\ypzc_{t}; \zpzc_{t}]\defeq \ln \ProbQ[\ypzc_{t},\zpzc_{t}]/\ProbQ[\ypzc_{t}]\ProbQ[\zpzc_{t}]$ is the bare mutual information between $\ypzc_{t}$ and $\zpzc_{t}$ and we use \eqnref{eq:PBc2} to derive the last equality.
From this, we can similarly obtain detailed, integral, and average FRs with information as follows:
\begin{align}
e^{-(\CMF_{0}[\ypzc_{t}] + i[\ypzc_{t}; \zpzc_{t}]-\CMF^{\cs}[\ypzc_{t},\zpzc_{t}])} =  \frac{\ProbP_{\LF,F}[\ypzc_{t}|\zpzc_{t}]\ProbQ[\zpzc_{t}]}{\ProbQ[\ypzc_{t},\zpzc_{t}]},\label{eq:DFC1}
\end{align}
\begin{align}
\average{e^{-(\CMF_{0}[\ypzc_{t}] + i[\ypzc_{t}; \zpzc_{t}]-\CMF^{\cs}[\ypzc_{t},\zpzc_{t}])}}_{\ProbQ[\ypzc_{t},\zpzc_{t}]} = 1,
\end{align}
and 
\begin{align}
\average{\CMF^{\cs}}_{\ProbQ} = & \average{\CMF_{0}}_{\ProbQ}+\MI^{\ypzc; \zpzc}  - \KLD_{\loss}^{\cs},\label{eq:AFRC1}
\end{align}
where $\MI^{\ypzc;\zpzc} \defeq \average{i[\ypzc_{t}; \zpzc_{t}]}_{\ProbQ[\ypzc_{t},\zpzc_{t}]}$ is the path-wise mutual information between $\ypzc_{t}$ and $\zpzc_{t}$, and 
\begin{align}
\KLD_{\loss}^{\cs} \defeq & \KL{\ProbP_{B}[\xpzc_{t},\ypzc_{t},\zpzc_{t}]}{\ProbP_{\LF}[\ypzc_{t}||\xpzc_{t}]\ProbP_{F}[\xpzc_{t}||\zpzc_{t}]\ProbQ[\zpzc_{t}]} \notag \\
=& \KL{\ProbQ[\ypzc_{t},\zpzc_{t}]}{\ProbP_{\LF,F}[\ypzc_{t}|\zpzc_{t}]\ProbQ[\zpzc_{t}]} \label{eq:Dsloss}.
\end{align}
The way information terms involved in Eqs. (\ref{eq:DFC1})-(\ref{eq:AFRC1}) is the same as the way those appearing in the Sagawa-Ueda relations, where the Maxwell demon and feedback regulation are involved\cite{Sagawa:2012wi}.
Because of the non-negativity of $\KLD_{\loss}^{\cs}$, $\average{\CMF^{\cs}}_{\ProbQ}$ is upper bounded by $\average{\CMF_{0}}_{\ProbQ}+\MI^{\ypzc; \zpzc}$:
\begin{align}
\average{\CMF^{\cs}}_{\ProbQ} \le & \average{\CMF_{0}}_{\ProbQ}+\MI^{\ypzc; \zpzc}=\average{\LF_{\max}}_{\ProbQ} - \mathcal{S}_{\ypzc|\zpzc}[\ProbQ],
\end{align}
where $\mathcal{S}_{\ypzc|\zpzc}[\ProbQ] \defeq - \average{\ln \ProbQ[\ypzc_{t}|\zpzc_{t}]}_{\ProbQ[\ypzc_{t},\zpzc_{t}]}$ is the conditional entropy of $\ProbQ[\ypzc_{t},\zpzc_{t}]$.
If the history of signal $\zpzc_{t}$ has perfect information on the history of $\ypzc_{t}$, the upper bound reaches $\average{\LF_{\max}}_{\ProbQ} $.
As in the bet-hediging situation, maximization of the average fitness $\average{\CMF^{\cs}}_{\ProbQ}$ with sensing is also dual to the minimization of the relative entropy $\KLD_{\loss}^{\cs}$:
\begin{align}
\{\ProbT_{F}^{*},\ProbT_{\LF}^{*}\} &\defeq\arg\max_{\{\ProbT_{F},\ProbT_{\LF}\}}\average{\CMF^{\cs}}_{\ProbQ}
=\arg \min_{\{\ProbT_{F},\ProbT_{\LF}\}}  \KLD_{\loss}^{\cs},\label{eq:optw}
\end{align}
where we use $*$ to denote the optimal $\ProbT_{F}$ and $\ProbT_{\LF}$ with sensing.
The duality indicates that $\average{\CMF^{\cs}}_{\ProbQ}$ achieves the bound $\average{\CMF_{0}}_{\ProbQ}+\MI^{\ypzc; \zpzc}$ only when $\ProbP^{*}_{\LF, F}[\ypzc_{t}|\zpzc_{t}]=\ProbQ[\ypzc_{t}|\zpzc_{t}]$ holds.
As in the bet-heding, if $\ProbP^{*}_{\LF, F}[\ypzc_{t}|\zpzc_{t}]=\ProbQ[\ypzc_{t}|\zpzc_{t}]$ holds, the time backward path probability $\ProbP_{B}^{\cs,*}[\xpzc_{t},\ypzc_{t},\zpzc_{t}]$ equals the time-forward path probability:
\begin{align}
\ProbP_{B}^{\cs,*}[\xpzc_{t},\ypzc_{t},\zpzc_{t}] = \ProbP_{\LF}^{*}[\ypzc_{t}||\xpzc_{t}]\ProbP_{F}^{*}[\xpzc_{t}||\zpzc_{t}]\ProbQ[\zpzc_{t}].\label{eq:CD_Common1}
\end{align}
Marginalization of this equation leads to the consistency condition $\ProbP_{F}^{*}[\xpzc_{t}||\zpzc_{t}]=\ProbP_{B}^{\cs,*}[\xpzc_{t}|\zpzc_{t}] ( \defeq \sum_{\ypzc_{t}}\ProbP_{B}^{\cs,*}[\xpzc_{t}|,\ypzc_{t},\zpzc_{t}]\ProbQ[\ypzc_{t}|\zpzc_{t}])$ derived in reference\cite{Kobayashi:2015eca}.
Moreover, $\ProbP_{\LF, F}[\ypzc_{t}|\zpzc_{t}]\ProbQ[\zpzc_{t}]$ is the conjugate environment and signal of a given pair of strategies $\{\ProbT_{F}, \ProbT_{\LF}\}$ under which it achieves the bound.
From the detailed FR in \eqnref{eq:DFC1}, we also see that the fitness of a given strategy can exceed the bound by chance as $\CMF^{\cs}[\ypzc_{t},\zpzc_{t}]>\CMF_{0}[\ypzc_{t}]+i[\ypzc_{t}; \zpzc_{t}]$ when the realized pair of environmental and signal histories $\{\ypzc_{t},\zpzc_{t}\}$ appears more frequently in the conjugate environment than in the actual environment as $\ProbP_{\LF, F}[\ypzc_{t}|\zpzc_{t}]\ProbQ[\zpzc_{t}]> \ProbQ[\ypzc_{t},\zpzc_{t}]$.

\subsection{Achievability of the bound and causality}
The necessary and sufficient condition $\ProbP^{*}_{\LF, F}[\ypzc_{t}|\zpzc_{t}]=\ProbQ[\ypzc_{t}|\zpzc_{t}]$ for achieving the bound means that the optimal metabolic allocation and phenotype switching strategy together implement the Bayesian computation of the posterior distribution $\ProbQ[\ypzc_{t}|\zpzc_{t}]$ of $\ypzc_{t}$ given the history of the sensed signal $\zpzc_{t}$.
Under the constraint that $\ProbP_{\LF, F}[\ypzc_{t}|\zpzc_{t}]$ satisfies a causal relation as $\ProbP_{\LF, F}[\ypzc_{t}|\zpzc_{t}]=\sum_{\xpzc_{t}}\ProbP_{\LF}[\ypzc_{t}||\xpzc_{t}]\ProbP_{F}[\xpzc_{t}||\zpzc_{t}]$,
 $\ProbP^{*}_{\LF, F}[\ypzc_{t}|\zpzc_{t}]=\ProbQ[\ypzc_{t}|\zpzc_{t}]$ cannot be realized in general because $\ProbQ[\ypzc_{t}|\zpzc_{t}]$ does not necessarily satisfy the causality relation between  $\ypzc_{t}$ and $\zpzc_{t}$.
If, for example, the metabolic allocation strategy is of Kelly's type as $\ProbT_{\LF}(y|x)=\delta_{y,x}$, the phenotypic switching strategy must satisfy
$\ProbP^{*}_{F}[\xpzc_{t}||\zpzc_{t}]=\ProbQ[\xpzc_{t}|\zpzc_{t}]$ to achieve the bound where $\ProbQ[\xpzc_{t}|\zpzc_{t}]\defeq\left.\ProbQ[\ypzc_{t}|\zpzc_{t}]\right|_{\ypzc_{t}=\xpzc_{t}}$.
By definition, $\ProbP_{F}[\xpzc_{t}||\zpzc_{t}]$ should satisfy the causal relation that $x(t)$ depends only on the past and/or current states of $z(t)$.
However, $\ProbQ[\ypzc_{t}|\zpzc_{t}]$ may not necessarily be causal because the conditioning by histories biases the past states of the conditioned history.
\com{For example, consider the case that the signal $\zpzc_{t}$ is causally generated from the environment $\ypzc_{t}$ and there is no feedback from $\zpzc_{t}$ to $\ypzc_{t}$ as in Fig.2(A) and (C).
If we observe $\zpzc_{t}=\{\cdots, z(t-2), z(t-1),z(t)\}=\{\cdots, \elz_{1},\elz_{1},\elz_{2}\}$, we may infer that the environmental state changes from $\ely_{1}$ to $\ely_{2}$ at time $t$.
However, if we further observe $\{z(t+1),z(t+2)\}=\{\elz_{1},\elz_{1}\}$, then we may change our prediction such that $y(t)=\ely_{1}$ and $z(t)=\elz_{2}$ was simply an error of the signal.
This intuitive observation illustrates that the inferred state of $y(t)$ can be affected by the future observation of the signal, e.g., $z(t+1)$ and $z(t+2)$.
For exactly the same reason, in $\ProbQ[\ypzc_{t}|\zpzc_{t}]$, the past state in the environmental history $\ypzc_{t}$ is affected by the future observations of the signal in $\zpzc_{t}$.
Because of the dependency on future state of the signal, $\ProbQ[\ypzc_{t}|\zpzc_{t}]$ cannot be represented causally in general.
}
An exception is that the environmental history is causally and memorylessly generated from the signal as $\ProbQ[\ypzc_{t}||\zpzc_{t}]=\prod_{\tau=0}^{t}\ProbT_{E}(y_{\tau}|z_{\tau})$, which is not realistic because we usually expect the signal to be generated from environment and not \textit{vice versa}.
Thus, 
\begin{align}
 \KLD_{\loss}^{\cs^{*}} \defeq \min_{\{\ProbT_{F},\ProbT_{\LF}\}} \KLD_{\loss}^{\cs},
 \end{align}
  is not generally zero, and  $\KLD_{\loss}^{\cs}$ contains not only the loss due to suboptimality of strategies, but also the loss from the causal constraints in exploiting the information of $\zpzc_{t}$ for phenotype switching.

\begin{figure}
\includegraphics[width=\linewidth]{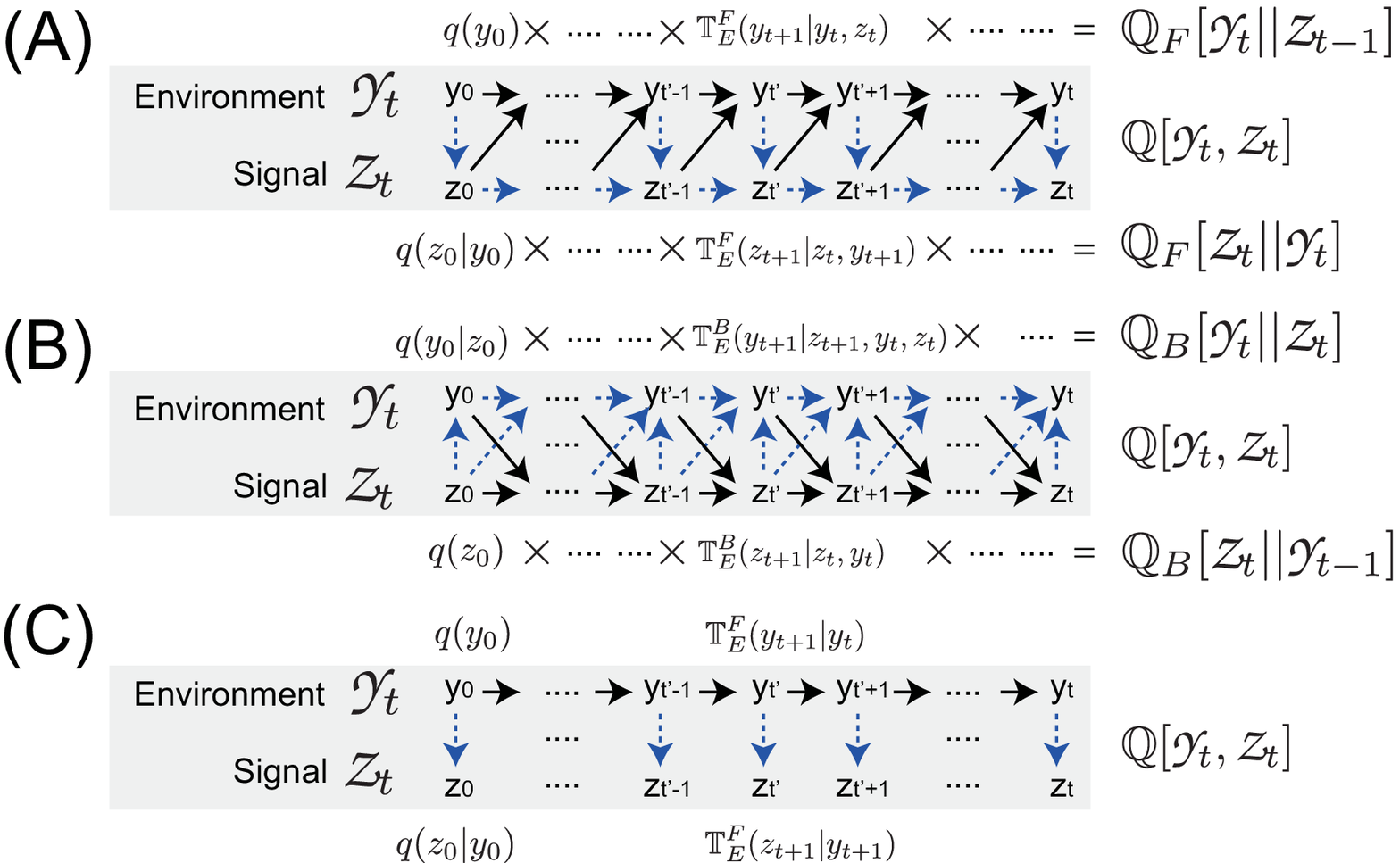}
\caption{\label{fig5} Causal dependency of environmental and signal histories. 
(A) A causal relation between environment and signal, in which the environment causally drives the signal.
Black solid arrows indicate that $y_{t+1}$ is affected by its one-step past state $y_{t}$ and that of the signal $z_{t}$. 
Blue dashed arrows represent that $z_{t+1}$ is affected by the current environment $y_{t+1}$ and the one-step past signal $z_{t}$.
(B) The causal decomposition of $\ProbQ[\ypzc_{t},\zpzc_{t}]$ with the reversed causal relation between the environment and the signal, in which the signal causally drives the environment.
Black solid arrows indicate that $z_{t+1}$ is affected by its one-step past environment $y_{t}$ and that of the signal $z_{t}$. 
Blue dashed arrows represent that $y_{t+1}$ is affected by the current and one-step past signal $z_{t+1}$ and $z_{t}$ and the one-step past environment $y_{t}$.
(C) The causal relation between the environment and the signal described in Figs. 2 (A) and (C), and used in the numerical simulation (Figs. 3 and 5).
}
\end{figure}

\section{Directed information and Causal FRs with sensing}\label{sec:CFRs}
By taking account of the causality relation between $\ypzc_{t}$ and $\zpzc_{t}$ in more detail, we can obtain tighter FRs, which illustrate the problem of the causality more explicitly.
Let us additionally assume that $\ProbQ[\ypzc_{t},\zpzc_{t}]$ has a causal and Markov relation as depicted  in Fig. 4 (A).
The relation is represented as 
\begin{align}
\ProbQ[\ypzc_{t},\zpzc_{t}] &=\left[\prod_{\tau=0}^{t-1}\ProbT_{E}^{F}(z_{\tau+1}|y_{\tau+1},z_{\tau})\ProbT_{E}^{F}(y_{\tau+1}|y_{\tau},z_{\tau})\right]q(y_{0},z_{0}),\notag \\
&\defeq \ProbQ_{F}[\zpzc_{t}||\ypzc_{t}]\ProbQ_{F}[\ypzc_{t}||\zpzc_{t-1}],\label{eq:Qcausal}
\end{align}
where $\ProbQ_{F}[\zpzc_{t}||\ypzc_{t}] \defeq \prod_{\tau=0}^{t-1}\ProbT_{E}^{F}(z_{\tau+1}|y_{\tau+1},z_{\tau})q(z_{0}|y_{0})$ and $\ProbQ_{F}[\ypzc_{t}||\zpzc_{t-1}] \defeq \prod_{\tau=0}^{t-1}\ProbT_{E}^{F}(y_{\tau+1}|y_{\tau},z_{\tau})q(y_{0})$.
Here, the signal $z(t)$ and environment $y(t)$ are Markovean and dependent  on the past states of the environment and the signal.
By applying Bayes' theorem, $\ProbQ[\ypzc_{t},\zpzc_{t}]$ admits another causal decomposition (Fig. 4 (B)) with reversed causality between $\ypzc_{t}$ and $\zpzc_{t}$: 
\begin{align}
\ProbQ[\ypzc_{t},\zpzc_{t}] = &\left[\prod_{t=0}^{t-1}\ProbT_{E}^{B}(y_{t+1}|z_{t+1}, y_{t}, z_{t})\right]\left[\prod_{t=0}^{t-1}\ProbT_{E}^{B}(z_{t+1}|y_{t},z_{t})\right],\notag \\
& \times q(y_{0},z_{0})\notag \\
\defeq & \ProbQ_{B}[\ypzc_{t}||\zpzc_{t}]\ProbQ_{B}[\zpzc_{t}||\ypzc_{t-1}] \label{eq:QdecombB},
\end{align}
where 
\begin{align}
&\ProbT_{E}^{B}(z_{t+1}|y_{t},z_{t})  \defeq \sum_{y_{t+1}}\ProbT_{E}^{F}(z_{t+1}|y_{t+1},z_{t})\ProbT_{E}^{F}(y_{t+1}|y_{t},z_{t}),\\
&\ProbT_{E}^{B}(y_{t+1}|z_{t+1}, y_{t}, z_{t}) \defeq \frac{\ProbT_{E}^{F}(z_{t+1}|y_{t+1},z_{t})\ProbT_{E}^{F}(y_{t+1}|y_{t},z_{t})}{\ProbT_{E}^{B}(z_{t+1}|y_{t},z_{t})},
\end{align}
$\ProbQ_{B}[\ypzc_{t}||\zpzc_{t}]\defeq \prod_{t=0}^{t-1}\ProbT_{E}^{B}(y_{t+1}|z_{t+1}, y_{t}, z_{t}) q(y_{0}|z_{0})$, and $\ProbQ_{B}[\zpzc_{t}||\ypzc_{t-1}]\defeq \prod_{t=0}^{t-1}\ProbT_{E}^{B}(z_{t+1}|y_{t},z_{t})q(z_{0})$.
Because of the meaning of Bayes' theorem, $\ProbT_{E}^{B}(y_{t+1}|z_{t+1}, y_{t}, z_{t})$ is interpreted as the posterior distribution of $y_{t+1}$ inferred from observations, $z_{t+1}$, $y_{t}$, and $z_{t}$.
Thus, $\ProbQ_{B}[\ypzc_{t}||\zpzc_{t}]$ is equivalent to the posterior path probability of the environment inferred by the sequential Bayesian inference of $y_{t+1}$ given the observation of the signal $z_{t}$ and the past state of the environment $y_{t}$.
Note that $\ProbT_{E}^{B}(y_{t+1}|z_{t+1}, y_{t}, z_{t})$ is slightly different from the usual sequential Bayesean inference to estimate the hidden state $y(t)$ only from observation $\zpzc_{t}$ in which we cannot use the past environmental state $y_{t}$ for inference.
Moreover, $\ProbQ_{B}[\ypzc_{t}||\zpzc_{t}]$ is also different from $\ProbQ_{B}[\ypzc_{t}|\zpzc_{t}]$ because the posterior path probability $\ProbQ_{B}[\ypzc_{t}|\zpzc_{t}]$ is computed after observing the whole history of $\zpzc_{t}$  rather than by conducting Bayesian inference sequentially as in $\ProbQ_{B}[\ypzc_{t}||\zpzc_{t}]$.

\Eqnref{eq:QdecombB} yields a tighter decomposition of the fitness:
\begin{align}
\CMF^{\cs}[\ypzc_{t},\zpzc_{t}] &= \CMF_{0}[\ypzc_{t}] + i[\zpzc_{t} \to \ypzc_{t}] \notag \\
& -\ln \frac{\ProbP_{B}^{\cs}[\xpzc_{t}, \ypzc_{t},\zpzc_{t}]}{\ProbP_{\LF}[\ypzc_{t}||\xpzc_{t}]\ProbP_{F}[\xpzc_{t}||\zpzc_{t}]\ProbQ_{B}[\zpzc_{t}||\ypzc_{t-1}]}.
\end{align}
where $i[\zpzc_{t} \to \ypzc_{t}]  \defeq \ln \ProbQ[\ypzc_{t},\zpzc_{t}]/\ProbQ_{B}[\zpzc_{t}||\ypzc_{t-1}]\ProbQ[\ypzc_{t}]$ is the bare directed information from $\zpzc_{t}$ to $\ypzc_{t}$\cite{Permuter:2011jra}.
This decomposition leads to the following detailed FR:
\begin{align}
&e^{-(\CMF_{0}[\ypzc_{t}]+i[\zpzc_{t} \to \ypzc_{t}]-\CMF^{\cs}[\ypzc_{t},\zpzc_{t}])} \notag \\
&\qquad \qquad  = \frac{\ProbP_{\LF}[\ypzc_{t}||\xpzc_{t}]\ProbP_{F}[\xpzc_{t}||\zpzc_{t}]\ProbQ_{B}[\zpzc_{t}||\ypzc_{t-1}]}{\ProbP_{B}^{\cs}[\xpzc_{t}, \ypzc_{t},\zpzc_{t}]},\notag \\
&\qquad \qquad  = \frac{\ProbP_{\LF,F}[\ypzc_{t}|\zpzc_{t}]\ProbQ_{B}[\zpzc_{t}||\ypzc_{t-1}]}{\ProbQ[\ypzc_{t},\zpzc_{t}]}.\label{eq:DFRcSens}
\end{align}
Because $\ProbP_{\LF}[\ypzc_{t}||\xpzc_{t}]\ProbP_{F}[\xpzc_{t}||\zpzc_{t}]\ProbQ_{B}[\zpzc_{t}||\ypzc_{t-1}]$ forms a joint path probability among $\xpzc_{t}$, $\ypzc_{t}$, and $\zpzc_{t}$, which satisfies $\sum_{\xpzc_{t},\ypzc_{t},\zpzc_{t}}\ProbP_{\LF}[\ypzc_{t}||\xpzc_{t}]\ProbP_{F}[\xpzc_{t}||\zpzc_{t}]\ProbQ_{B}[\zpzc_{t}||\ypzc_{t-1}]=1$, we obtain the integral and average FRs as
\begin{align}
\average{e^{-(\CMF_{0}[\ypzc_{t}]+i[\zpzc_{t} \to \ypzc_{t}]-\CMF^{\cs}[\ypzc_{t},\zpzc_{t}])}}_{\ProbQ[\ypzc_{t},\zpzc_{t}]} =1, \label{eq:IFRcSens}
\end{align}
and 
\begin{align}
\average{\CMF^{\cs}}_{\ProbQ}=\average{\CMF_{0}}_{\ProbQ}+\MI^{\zpzc_{t} \to \ypzc_{t}}- \KLD_{\loss}^{\cs, d} \le  \average{\CMF_{0}}_{\ProbQ}+\MI^{\zpzc_{t} \to \ypzc_{t}},\label{eq:AFRcSens}
\end{align}
where 
\begin{align}
\KLD_{\loss}^{\cs,d} &\defeq \KL{\ProbP_{B}[\xpzc_{t},\ypzc_{t},\zpzc_{t}]}{\ProbP_{\LF}[\ypzc_{t}||\xpzc_{t}]\ProbP_{F}[\xpzc_{t}||\zpzc_{t}]\ProbQ_{B}[\zpzc_{t}||\ypzc_{t-1}]},\notag \\
& = \KL{\ProbQ[\ypzc_{t},\zpzc_{t}]}{\ProbP_{\LF,F}[\ypzc_{t}|\zpzc_{t}]\ProbQ_{B}[\zpzc_{t}||\ypzc_{t-1}]}
\end{align}
 and $\MI^{\zpzc_{t} \to \ypzc_{t}} \defeq \KL{\ProbQ[\ypzc_{t},\zpzc_{t}]}{\ProbQ_{B}[\zpzc_{t}||\ypzc_{t-1}]\ProbQ[\ypzc_{t}]}$.
$\MI^{\zpzc_{t} \to \ypzc_{t}}$ is the directed information that quantifies the amount of information $\zpzc_{t}$ has for inferring $\ypzc_{t}$\cite{Permuter:2011jra}. 
If $\ProbQ_{B}[\ypzc_{t}||\zpzc_{t}]=\ProbQ[\ypzc_{t}]$, $\MI^{\zpzc_{t} \to \ypzc_{t}}=0$ as well as $i[\zpzc_{t} \to \ypzc_{t}]=0$ hold, which means that $\zpzc_{t}$ is useless for inferring $\ypzc_{t}$ causally.
$\average{\CMF_{0}}_{\ProbQ}+\MI^{\zpzc_{t} \to \ypzc_{t}}$ in \eqnref{eq:AFRcSens} is a tighter bound for the average fitness with sensing because 
\begin{align}
\MI^{\ypzc; \zpzc} &= \average{\ln\frac{\ProbQ[\ypzc_{t},\zpzc_{t}]}{\ProbQ_{B}[\zpzc_{t}||\ypzc_{t-1}]\ProbQ[\ypzc_{t}]}+\ln\frac{\ProbQ[\ypzc_{t},\zpzc_{t}]}{\ProbQ_{B}[\ypzc_{t}||\zpzc_{t}]\ProbQ[\zpzc_{t}]}}_{\ProbQ[\ypzc_{t},\zpzc_{t}]}\notag \\
&= \MI^{\zpzc_{t} \to \ypzc_{t}} + \MI^{\ypzc_{t} \to \zpzc_{t}} \ge \MI^{\zpzc_{t} \to \ypzc_{t}}
\end{align}
holds, where $\MI^{\ypzc_{t} \to \zpzc_{t}} \defeq \KL{\ProbQ[\ypzc_{t},\zpzc_{t}]}{\ProbQ_{B}[\ypzc_{t}||\zpzc_{t}]\ProbQ[\zpzc_{t}]}$ \com{is the directed information from $\ypzc_{t}$ to $\zpzc_{t}$}.
Moreover, $\MI^{\ypzc_{t} \to \zpzc_{t}}$ is the residual information in $\MI^{\ypzc; \zpzc}$ that has no fitness value under the causal constraints.

\subsection{Numerical verification of FRs with sensing}
To demonstrate the validity of FRs in Eqs. (\ref{eq:DFRcSens})-(\ref{eq:AFRcSens}), we conduct a modified numerical simulation of the population dynamics to incorporate sensing (see also Appendix \ref{ap:num}). 
The stochastic laws of environmental dynamics $\ProbT_{E}^{F}(y_{t+1}|y_{t},z_{t})$ (Fig.2 (A)) and the metabolic allocation strategy $\ProbT_{\LF}(y_{t}|x_{t})$ (Fig.2 (B)) are the same as those of the simulations in Fig. 3. 
The sensing signal is assumed to have two states, $\elz_{1}$ and $\elz_{2}$ (Fig. 2 (C)). 
If the environment is either in $\ely_{1}$ or in $\ely_{2}$, the signal gives the corresponding $\elz_{1}$ and $\elz_{2}$ with  \com{$90\%$} accuracy. If the environment is in $\ely_{3}$, the signal produces either $\elz_{1}$ or $\elz_{2}$ with eqaul probability (Fig. 2 (C)). 
Because the signal is memorylessly generated from the environment at every time point, the environment and signal  together form a Markov relation as depicted in Fig. 4 (C).
By employing the sensed signal, the organisms switch to the phenotypic state $\elx_{1}$ with a $95 \%$ chance when the sensed signal is $\elz_{1}$ and to $\elx_{2}$ when the signal is $\elz_{2}$ as shown in Fig. 2 (D).

Figures 5 (A) and (B) show two trajectories of the population size under the same realizations of environment as in Figs. 3 (A) and (B), respectively.
In Fig. 5 (A), the total population size $\popN_{t}^{\ypzc,\zpzc}=e^{\CMF^{\cs}[\ypzc_{t},\zpzc_{t}]}$ (dashed line) is less than $e^{\CMF_{0}[\ypzc_{t}]+i[\zpzc_{t} \to \ypzc_{t}]}$ (red line) whereas $\popN_{t}^{\ypzc,\zpzc}$ frequently becomes greater than $e^{\CMF_{0}[\ypzc_{t}]+i[\zpzc_{t} \to \ypzc_{t}]}$ in Fig. 5 (B).
Even with sensing, both $\CMF^{\cs}[\ypzc_{t},\zpzc_{t}]$ and $\CMF_{0}[\ypzc_{t}]+i[\zpzc_{t} \to \ypzc_{t}]$ fluctuate substantially as shown in Figs. 5 (C) and (D) depending on the realization of the environment and signal. 
Due to this fluctuation, $\CMF^{\cs}[\ypzc_{t},\zpzc_{t}]$ sometimes become greater than $e^{\CMF_{0}[\ypzc_{t}]+i[\zpzc_{t} \to \ypzc_{t}]}$ as in Fig. 5 (E) to satisfy the integral FR (\ref{eq:IFRcSens}) as demonstrated in Fig. 5 (F).

\begin{figure}[H]
\includegraphics[width=\linewidth]{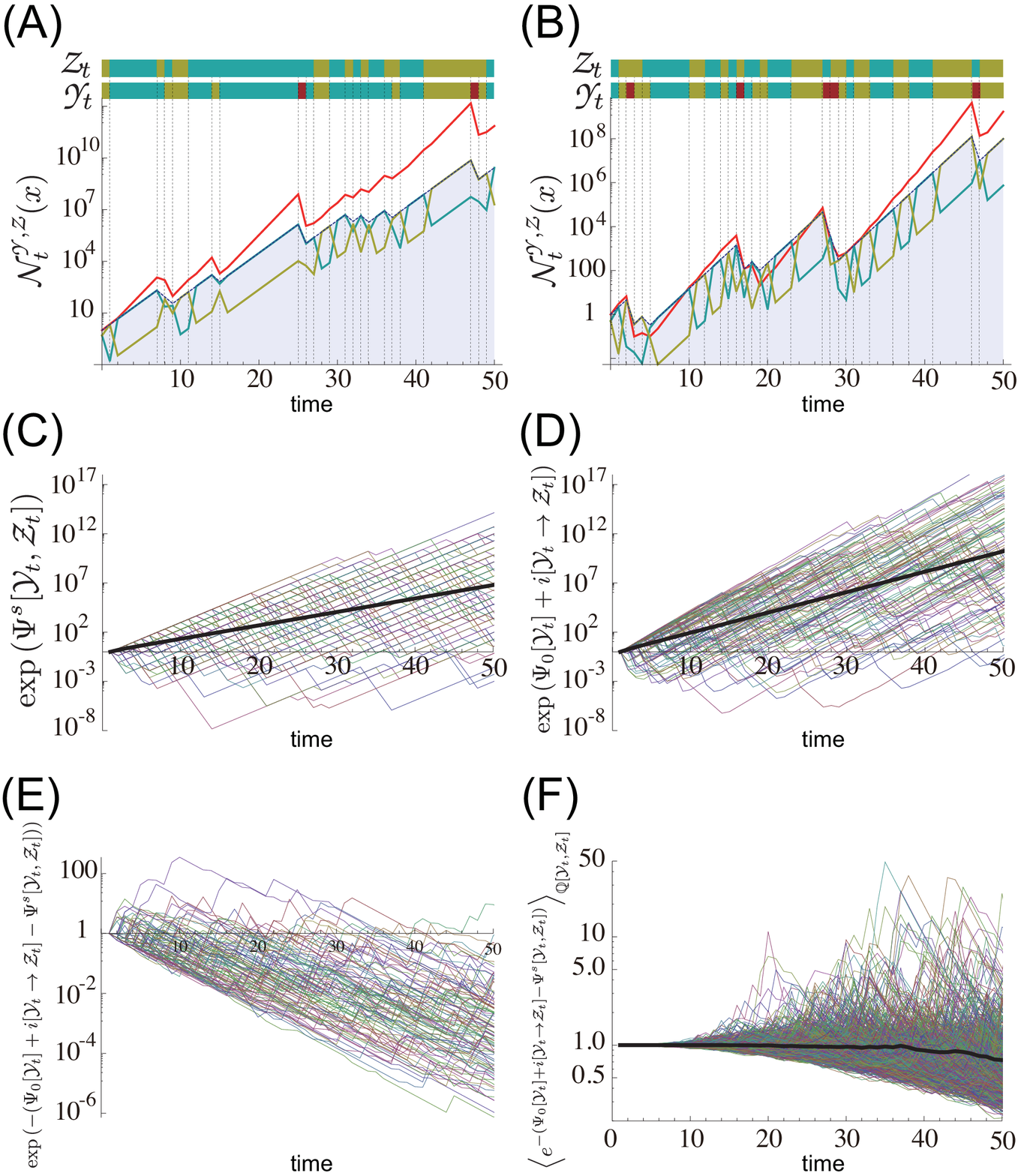}
\caption{\label{fig4}
 Numerical simulation of the population dynamics with sensing defined in Fig. 2.
(A, B) Two independent sample histories of environment $\ypzc_{t}$ and signal $\zpzc_{t}$. 
The dynamic of $\popN_{t}^{\ypzc,\zpzc}(x_{t})$, $\popN_{t}^{\ypzc,\zpzc}=e^{\CMF^{s}[\ypzc_{t},\zpzc_{t}]}$, and $e^{\CMF^{0}[\ypzc_{t}]+i[\zpzc_{t} \to \ypzc_{t}]}$ under the two realizations of the environmental and signal histories \com{obtained by solving \eqnref{eq:popdyn}}. 
The color bars on the graphs represent the states of the environment and the signal at each time point. 
The correspondence of the colors and environmental and signal states is as shown in Fig. 2 (A) and (C). 
Cyan and yellow lines represent $\popN_{t}^{\ypzc,\zpzc}(\elx_{1})$ and $\popN_{t}^{\ypzc,\zpzc}(\elx_{2})$, respectively.
The dashed blue line with filled grey style is  $\popN_{t}^{\ypzc,\zpzc}$. 
The red line is $e^{\CMF_{0}[\ypzc]+i[\zpzc_{t} \to \ypzc_{t}]}$.
(C) Stochastic behaviors of population fitness $\CMF^{s}[\ypzc_{t},\zpzc_{t}]$ for $100$ independent samples of environmental and signal histories. 
Each colored thin line represents $e^{\CMF^{s}[\ypzc_{t},\zpzc_{t}]}=\popN_{t}^{\ypzc,\zpzc}/\popN_{0}^{\ypzc,\zpzc}$ for each realization of the environmental and signal histories. 
The thick black line is $e^{\average{\CMF^{s}[\ypzc_{t},\zpzc_{t}]}_{\ProbQ[\ypzc_{t},\zpzc_{t}]}}$.
(D) Stochastic behaviors of $\CMF_{0}[\ypzc_{t}]+i[\zpzc_{t} \to \ypzc_{t}]$  for the same $100$ samples of environmental and signal histories as those in (C). 
Each colored thin line represents $e^{\CMF_{0}[\ypzc_{t}]+i[\zpzc_{t} \to \ypzc_{t}]}$ for each realization of the same environmental and signal histories as in (C). 
(E) Stochastic behaviors of $e^{-(\CMF_{0}[\ypzc_{t}]+i[\zpzc_{t} \to \ypzc_{t}]-\CMF^{s}[\ypzc_{t}])}$ for the same $100$ samples of environmental and signal histories as those in (C). 
Each colored thin line represents $e^{-(\CMF_{0}[\ypzc_{t}]+i[\zpzc_{t} \to \ypzc_{t}]-\CMF^{s}[\ypzc_{t}])}$ for each realization of the same environmental and signal histories as in (C).
(F) $\average{e^{-(\CMF_{0}[\ypzc_{t}]+i[\zpzc_{t} \to \ypzc_{t}]-\CMF^{s}[\ypzc_{t}])}}_{\ProbQ[\ypzc_{t},\zpzc_{t}]}$ calculated empirically by the numerical simulations of sample paths of $e^{-(\CMF_{0}[\ypzc_{t}]+i[\zpzc_{t} \to \ypzc_{t}]-\CMF^{s}[\ypzc_{t}])}$.
The meaning of the thick and thin lines are the same as in Fig. 3.
}
\end{figure}

\subsection{Achievability of the bound}
The bound in \eqnref{eq:AFRcSens} can be achieved when $\ProbP_{\LF, F}^{*}[\ypzc_{t}|\zpzc_{t}]=\ProbQ_{B}[\ypzc_{t}||\zpzc_{t}]$ holds, which also means that $\ProbQ^{*}[\ypzc_{t},\zpzc_{t}]\defeq \ProbQ^{*B}[\ypzc_{t}||\zpzc_{t}]\ProbQ_{B}[\zpzc_{t}||\ypzc_{t-1}]$ is the conjugate environment for a given $\{\ProbT_{F},\ProbT_{\LF}\}$ where $\ProbQ^{*B}[\ypzc_{t}||\zpzc_{t}]=\ProbP_{\LF, F}[\ypzc_{t}|\zpzc_{t}]$.
Because $\ProbQ_{B}[\ypzc_{t}||\zpzc_{t}]$ is the path obtained by a type of sequential Bayesian inference $\ProbT_{E}^{B}(y_{t+1}|z_{t+1}, y_{t}, z_{t}) $, 
the bound is attained if phenotype switching with sensing and metabolic allocation, as a whole, implement the same Bayesian computation as $\ProbT_{E}^{B}(y_{t+1}|z_{t+1}, y_{t}, z_{t})$.
However, to estimate $y_{t+1}$, the Bayesian inference represented by $\ProbT_{E}^{B}(y_{t+1}|z_{t+1}, y_{t}, z_{t}) $ uses not only $z_{t+1}$ and $z_{t}$ but also the error-less information on the past environmental state $y_{t}$.
If organisms do not have any way to obtain perfect information on the past environment by sensing as they do not for biologically realistic situations, 
the metabolic allocation strategy should be of Kelly's type as $\ProbT_{\LF}^{*}(y_{t}|x_{t})=\delta_{y_{t},x_{t}}$ to achieve the bound without additional constraints on the environment.
With Kelly's strategy, the organisms can effectively obtain information on the past environment because the past phenotypic state $x(t)$ of the survived organisms becomes the same as the past environmental state $y(t)$ under this situation.
The optimal phenotype switching with Kelly's strategy for metabolic allocation should be 
\begin{align}
\ProbT_{F}^{*}(x_{t+1}|x_{t},z_{t+1},z_{t})=\left.\ProbT_{E}^{B}(y_{t+1}|z_{t+1}, y_{t}, z_{t})\right|_{\substack{y_{t+1}=x_{t+1}, \\y_{t}=x_{t}}}.\label{eq:coptKelly}
\end{align}
to attain the bound.
Another situation in which the bound can be achieved is when the environmental state $y_{t+1}$ depends not on $y_{t}$ but only on $z_{t}$ as $\ProbT_{E}(y_{t+1}|y_{t},z_{t})=\ProbT_{E}(y_{t+1}|z_{t})$.
We then have
\begin{align}
\ProbT_{E}^{B}(z_{t+1}|z_{t}) & = \sum_{y_{t+1}}\ProbT_{E}^{F}(z_{t+1}|y_{t+1})\ProbT_{E}^{F}(y_{t+1}|z_{t}),\\
\ProbT_{E}^{B}(y_{t+1}|z_{t+1}, y_{t}, z_{t}) &= \frac{\ProbT_{E}^{F}(z_{t+1}|y_{t+1})\ProbT_{E}^{F}(y_{t+1}|z_{t})}{\ProbT_{E}^{F}(z_{t+1}|z_{t})},
\end{align}
where $\ProbT_{E}^{B}(y_{t+1}|z_{t+1}, y_{t}, z_{t})$ is reduced to be independent of  $y_{t}$ as $\ProbT_{E}^{B}(y_{t+1}|z_{t+1}, y_{t}, z_{t})=\ProbT_{E}^{B}(y_{t+1}|z_{t+1}, z_{t})$.
Thus, the bound can be attained if the phenotypic switching and metabolic allocation strategies satisfy
\[
\sum_{x_{t}}\ProbT_{\LF}^{*}(y_{t+1}|x_{t+1})\ProbT_{F}^{*}(x_{t+1}|z_{t+1},z_{t})=\ProbT_{E}^{B}(y_{t+1}|z_{t+1}, z_{t}).
\]
Note that the phenotypic switching need not be dependent on the past phenotypic state in this situation because $z(t+1)$ and $z(t)$ contain all the relevant information on the future state of $y(t+1)$.

For both cases, the achievability of the bound is directly linked to the accessibility of the organisms to the past information that directly drives the current environmental state.
In addition, even when the bound is not achieved, $\KLD_{\loss}^{\cs,d^{*}}$ explicitly represents the loss of fitness associated with  the inaccessibility.

\subsection{Losses due to limited capacity and suboptimality}\label{sec:MMIs}
For general metabolic allocation strategies other than Kelly's type,  the phenotypic states of the survived organisms contain only imperfect information on the past environmental history.
In addition, the environment and signal may not always form the circular dependency $\dots \to y_{t} \to z_{t} \to y_{t+1} \to z_{t+1} \to \dots$.
In such situations, the loss is most generally represented by $\KLD_{\loss}^{\cs}$, which quantifies the difference between the time-backward path probability $\ProbP_{B}[\xpzc_{t}, \ypzc_{t}, \zpzc_{t}]$ and the time-forward path probability $\ProbP_{\LF}[\ypzc_{t}||\xpzc_{t}]\ProbP_{F}[\xpzc_{t}||\zpzc_{t}]\ProbQ[\zpzc_{t}]$.
The loss from different sources can be represented by using the following dissection of the fitness:
\begin{align}
\CMF^{\cs}[\ypzc_{t},\zpzc_{t}] &= \CMF_{0}[\ypzc_{t}] + i[\ypzc_{t}, \zpzc_{t}] -i_{B}^{\cs}[\ypzc_{t}, \zpzc_{t}|\xpzc_{t}] \\
& -\ln \frac{\ProbP_{B}^{\cs}[\ypzc_{t}|\xpzc_{t}]}{\ProbP_{\LF}[\ypzc_{t}||\xpzc_{t}]}- \ln\frac{\ProbP_{B}^{\cs}[\xpzc_{t}|\zpzc_{t}]}{\ProbP_{F}[\xpzc_{t}||\zpzc_{t}]},\label{eq:decomRetC}
\end{align}
where $i_{B}^{\cs}[\ypzc_{t}, \zpzc_{t}|\xpzc_{t}]  \defeq \ln \ProbP_{B}^{\cs}[\ypzc_{t},\zpzc_{t}|\xpzc_{t}]/\ProbP_{B}^{\cs}[\ypzc_{t}|\xpzc_{t}]\ProbP_{B}^{\cs}[\zpzc_{t}|\xpzc_{t}]$ is the bare conditional backward mutual information.
Although this representation does not immediately admit detailed and integral FRs, we can obtain the following decomposition of $\KLD_{\loss}^{\cs}$ into three terms:
\[
\KLD_{\loss}^{\cs}= \MI_{B}^{\ypzc; \zpzc|\xpzc} + \KLD_{\loss}^{\cs,\LF} + \KLD_{\loss}^{\cs,F},
\]
where 
\begin{align*}
\MI_{B}^{\ypzc; \zpzc|\xpzc} \defeq & \average{i_{B}^{\cs}[\ypzc_{t}, \zpzc_{t}|\xpzc_{t}]}_{\ProbP_{B}[\xpzc_{t},\ypzc_{t},\zpzc_{t}]},\\
\KLD_{\loss}^{\cs,\LF}\defeq &\average{\KL{\ProbP_{B}^{\cs}[\ypzc_{t}|\xpzc_{t}]}{\ProbP_{\LF}[\ypzc_{t}||\xpzc_{t}]}}_{\ProbP_{B}^{\cs}[\xpzc_{t}]}, \\
\KLD_{\loss}^{\cs,F}\defeq &\average{\KL{\ProbP_{B}^{\cs}[\xpzc_{t}|\zpzc_{t}]}{\ProbP_{F}[\xpzc_{t}||\zpzc_{t}]}}_{\ProbQ[\zpzc_{t}]}.
\end{align*}
$\MI_{B}^{\ypzc; \zpzc|\xpzc}$ is the residual information on the environment that the signal still has even if we know the retrospective history of the phenotype. 
Because the retrospective correlation between the phenotype and environment is induced by the selection, this residual information represents the information that is not used in the selection.
If the signal contains certain information on the environmental fluctuation that is nothing to do with the replication of the organisms, for example, the information on the existence of unmetabolizable artificial molecules, such information cannot be exploited in the phenotypic switching to choose a better phenotype for the current environmental state.
Thus, $\MI_{B}^{\ypzc; \zpzc|\xpzc}$ measures the amount of such useless information for fitness that the sensing signal conveys.
$\KLD_{\loss}^{\cs,\LF}$ accounts for the imperfectness of the metabolic allocation strategy, and
$\KLD_{\loss}^{\cs,F}$ quantifies the imperfectness of phenotype switching due to the causal constraint and suboptimality.

For example, in the case of the causally optimal strategy under Kelly's metabolic allocation in \eqnref{eq:coptKelly} with the causal structure in $\ProbQ[\ypzc_{t},\zpzc_{t}]$ (\eqnref{eq:Qcausal}), these quantities are reduced to
\begin{align*}
\MI_{B}^{\ypzc; \zpzc|\xpzc} = 0, \qquad \KLD_{\loss}^{\cs,\LF} =0, \qquad \KLD_{\loss}^{\cs,F} =\MI^{\ypzc_{t} \to \zpzc_{t}}.
\end{align*}
The first equation, $\MI_{B}^{\ypzc; \zpzc|\xpzc} = 0$, holds because organisms with the metabolic allocation strategy of Kelly's type can survive only when their phenotypic history is identical to the actual environmental history, and the retrospective phenotypic history $\xpzc_{t}$ contains perfect information on $\ypzc_{t}$. Thus, the residual $\MI_{B}^{\ypzc; \zpzc|\xpzc}$ becomes $0$.
Similarly, $\KLD_{\loss}^{\cs,\LF} =0$ because only organisms with their phenotypic history identical to their environmental hisotry can survive, and $\ProbP_{B}^{\cs}[\ypzc_{t}|\xpzc_{t}] = \delta_{\xpzc_{t},\ypzc_{t}}$ holds for Kelly's strategy. 
In contrast, $\KLD_{\loss}^{\cs,F}$ cannot be zero because the phenotypic switching strategy is causally constrained. The minimum loss due to this constraint is the directed information $\MI^{\ypzc_{t} \to \zpzc_{t}}$ that measures the causally useless information in $\MI^{\ypzc; \zpzc}$.

In a general situation, these three quantities are mutually related. 
For example, the amount of useless information for fitness $\MI_{B}^{\ypzc; \zpzc|\xpzc}$ depends on the choice of strategies.  
If phenotype switching does not use any information obtained from $z(t)$, then $\MI_{B}^{\ypzc; \zpzc|\xpzc}$ becomes  $\MI_{B}^{\ypzc; \zpzc}$.
They are appropriate information-theoretic quantities that account for the irrelevance of sensing and  the imperfectness of metabolic allocation and phenotypic switching for exploiting the sensed information.


\section{Summary and Discussion}\label{sec:Sum}
In this paper, we clarified the stochastic and information thermodynamic structures in population dynamics with and without sensing, to derive the bound of fitness and the fitness gain by sensing, in the form of FRs.
The detailed and integral FRs obtained substantially generalized the previous results on the average fitness value of information by showing that not only average but also the fluctuation of fitness is generically constrained as is the entropy production.
Such constraints manifest the possibility and the condition that the fitness of organisms with suboptimal strategies can be greater than that of the optimal one by chance due to environmental fluctuations, just as there is a finite probability of observing reversal of heat flow in a small thermal system.
Such rare events can be regarded as a violation of ESS because they may induce the takeover of the optimal organism by a suboptimal one in a finite population.
Nevertheless, the violation is somehow ruled to follow the integral and detailed FRs.
Moreover, the directed information is derived to be the tighter measure of the fitness value of information, in which the causal structure in the problem is explicitly accounted for.
The condition for achieving this bound is shown to be related to the ability of the organisms to conduct or implement a type of Bayesian sequential inference from the sensed information.
Finally, we derived three quantities that can measure the irrelevance of sensing and  the imperfectness of metabolic allocation and phenotypic switching for exploiting the sensed information.

All these results and generalizations are derived by employing the path-wise and the time-backward representation of population dynamics.
Among others, pivotal is the duality between the maximization of the average fitness and the minimization of the difference between the time-forward and time-backward path probabilities (eqs. (\ref{eq:optb}) and (\ref{eq:optw})).
The minimum loss of the average fitness due to causality is clearly described in this dual problem as
the deviation between the causal time-forward path probability and the non-causal time-backward path probability (\eqnref{eq:Dsloss}).
We believe that the path-wise and retrospective formulation of the population dynamics also play indispensable roles in  addressing other biologically relevant problems.

\subsection{Learning rules}
In this paper as well as in most of the previous works, the processes of attaining better strategies by mutation or learning  are rarely considered directly and explicitly\cite{Beaumont:2009gxa,Lambert:2014gc},  except in a reference\cite{Xue:2016fy}.
As we have demonstrated, the average optimal strategy may not dominate a population all the time in a fluctuating environment because of a rare event: the takeover by suboptimal strategies. 
Thus, the organisms and their corresponding strategies are expected to change non-stationarily over time.
In such a situation, an advantage may be gained by organisms that acquire an ability to adaptively change and learn the strategies as our brain and immune system do\cite{Mayer:2016jm}.
However, the fitness to be increased by adaptation is a population-level quantity whereas adaptation is conducted at the individual-level. 
The duality might be used to resolve this problem. 
As demonstrated in Kelly's strategy in the most extreme fashion, the survived organisms can retrospectively obtain certain information on the environment they experienced. 
As shown recently\cite{Xue:2016fy}, such information can be used to adjust time-forward strategies to switch phenotype and to allocate resource so as to reduce the discrepancy between time-forward and backward phenotypic histories rather than directly increasing the population fitness.
By employing our path-wise formulation, we may obtain more general results on the adaptive learning of strategies and its evolutionary advantages.

\subsection{Individual sensing}
Another problem that should be addressed is the type of sensing. 
In this paper, the sensing signal is treated as an extrinsic factor by assuming that all organisms have the same sensing signal.
Such an assumption is valid only when the sensing noise of the organisms is negligibly small.
A more general and biologically realistic situation is the individual sensing in which each organism receives different sensing signals owing to the sensing noise intrinsic to the organism\cite{Perkins:2009cg,BenJacob:2010ii,Kobayashi:2012ji,Brennan:2012cj}.
The fitness value of individual sensing has been rarely addressed, with the exception of a pioneering work\cite{Rivoire:2011fy}, which shows that the fitness value of individual sensing can be greater than the mutual information under restricted situations\cite{Rivoire:2011fy}.
A generalization of this result may be achieved by using our pat-hwise and retrospective formulation.

\subsection{Thermodynamics and evolution}
As we have clarified, thermodynamics and adaptation share the same fundamental mathematical structure.
Nevertheless, most attempts to bridge thermodynamics and adaptation or evolution, including ours, are just formal in the sense that the similarity shown is at the level of mathematics \cite{Lotka:1922ux,Lotka:1922wr,Iwasa:1988ux,deVladar:2011kz,Qian:2014ha}. 
However, an actual thermodynamics underlies the processes of replication and sensing of organisms\cite{England:2013ed,England:2015hl,Pugatch:2015gx}.
The thermodynamics must constrain the rate of replication and the efficiency of sensing, and the latter was investigated intensively in the context of stochastic and information thermodynamics, recently\cite{Barato:2014df,Govern:2014ef,Wolde:2016ih}.
Our relation between fitness and information must be consistent with the constraints imposed by the thermodynamics of these processes. 
Integration of the relation between entropy and information with that between fitness and information would be an indispensable step toward establishing the real thermodynamics of adaptation and evolution.

\begin{acknowledgments}
We acknowledge Yuichi Wakamoto, Takashi Nozoe, and Yoichiro Takahashi for useful discussions. 
This research is supported partially by an JST PRESTO program, JSPS KAKENHI 16H06155 and 16K17763, 
the Platform for Dynamic Approaches to Living System from MEXT and AMED, Japan, and the 2016 Inamori Research Grants Program, Japan.
\end{acknowledgments}

\appendix

\section{Decomposition of replication rate}\label{ap:decomp}
Let us consider a given replication rate $\lf(x,y)$. 
$e^{\lf(x,y)}$ determines a $(\# \setx \times \# \sety)$ matrix, and we define its row vectors by $\{\mathbf{v}_{x} \in (\mathbb{R}_{\ge0})^{\# \sety} | v_{x,y}=e^{\lf(x,y)},\, x \in \setx \}$.
When the number of the environmental states is equal to or greater than the phenotypic ones as $\# \sety \ge \# \setx$, 
$\{\mathbf{v}_{x}|x \in \setx\}$ can form a hyperplane, and we can fine a vector $\mathbf{u}\in (\mathbb{R}_{\ge0})^{\# \sety}$ that is orthogonal to the plane.
Let $\mathbf{q}_{0}$ be the normalization of $\mathbf{u}$ as $\mathbf{q}_{0}=\mathbf{u}/\sum_{y}u(y)$. 
By definition, $\mathbf{q}_{0}$ satisfies $ \mathbf{v}_{x} \cdot \mathbf{q}_{0}=\sum_{y}e^{\lf(x,y)}q_{0}(y) = \phi_{0}$ for all $x \in \setx$ where $\phi_{0}>0$ is a constant. 
This means that we can define a conditional probability $\ProbT_{\LF}(y|x)$ as
\begin{align}
\ProbT_{\LF}(y|x) \defeq \frac{e^{\lf(x,y)}q_{0}(y)}{\phi_{0}}.
\end{align}
Thus, we have the decomposition of $e^{\lf(x,y)}$ in \eqnref{eq:ek} and its variant used in \eqnref{eq:DecombK} as
\begin{align}
e^{\lf(x,y)} = \frac{\phi_{0}\ProbT_{\LF}(y|x)}{q_{0}(y)} = e^{\lf_{\max}(y)}\ProbT_{\LF}(y|x),\label{eq:decompK}
\end{align}
where $e^{\lf_{\max}(y)} \defeq \phi_{0}/q_{0}(y)$.
Thus, when the environmental states are more complex than the phenotypic ones as $\# \sety \ge \# \setx$, the decomposition of the replication rate, \eqnref{eq:ek}, is general enough even if we do not explicitly assume the relation between metabolic allocation strategy and replication rate as in \eqnref{eq:ek}.

If $\# \sety < \# \setx$, the decomposition does not necessarily exist.
Such a situation may occur when an organism has redundant phenotypic states, a fraction of which can be effectively realized by  linear combinations of the others.
While the redundant phenotypic states can appear in the process of mutation, $\# \sety < \# \setx$ is not biologically realistic because the environment is generally more complex than phenotype.

Finally, we note that the decomposition, \eqnref{eq:decompK}, is not unique except when $\# \setx = \# \sety$ and $\{\mathbf{v}_{x}|x \in \setx\}$ are linearly independent each others.
Nevertheless, our FRs hold irrespective of the choice of the decomposition.
Achievability of the maximum average fitness, \eqnref{eq:AFRbBound}, is affected by the choice because $\Phi_{0}$ depends on the decomposition and because $\ProbT(y|x)$ must be changed so that it satisfies \eqnref{eq:decompK} for a given $\lf(x,y)$.
While we apparently have a freedom to choose the metabolic allocation strategy, $\ProbT_{\LF}(y|x)$, for maximization of $\average{\CMF^{b}}_{\ProbQ}$ in \eqnref{eq:AFRbBound}, \eqnref{eq:AFRbBound} is virtually reduced to the problem of the maximization of the average fitness by changing only phenotypic switching strategy $\ProbP_{F}$ as
\begin{align}
\max_{\ProbT_{F}}\average{\CMF^{b}}_{\ProbQ}  \le \average{\CMF_{0}}_{\ProbQ},
\end{align}
because $\CMF^{b}$ is independent of the choice of the decomposition of $e^{\lf(x,y)}$.
This is the problem addressed in our previous work\cite{Kobayashi:2015eca}.
Tight upper bound of $\max_{\ProbT_{F}}\average{\CMF^{b}}_{\ProbQ}$ can be obtained by minimization of $\average{\CMF_{0}}_{\ProbQ}$ as 
\begin{align}
\max_{\ProbT_{F}}\average{\CMF^{b}}_{\ProbQ}  \le \min_{\ProbT_{\LF}}\average{\CMF_{0}}_{\ProbQ}.
\end{align}
We again stress that our integral and detailed FRs generally hold for any given $\ProbT_{F}$ and $\ProbT_{\LF}$.


\section{Numerical verification of FRs}\label{ap:num}
For the numerical simulations shown in Figs 3 and 5, we consider the case that the environmental dynamics is Markovean as $\ProbT_{E}^{F}(y_{t+1}|y_{t})$, and that the signal is memorylessly generated from the environment as $\ProbT_{E}^{F}(z_{t+1}|z_{t+1})$.

Realizations of the environmental history $\ypzc_{t}$ and the sensing signal $\zpzc_{t}$ are obtained by conducting a finite-state Markov transition by following $\ProbT_{E}^{F}(y_{t+1}|y_{t})$ and $\ProbT_{E}^{F}(z_{t+1}|y_{t+1})$.
For a given pair of realizations $\{\ypzc_{t},\zpzc_{t}\}$, the fitnesses $\CMF^{b}[\ypzc_{t}]$ (without sensing) and $\CMF^{\cs}[\ypzc_{t}]$ (with sensing) are calculated recursively as 
\begin{align*}
\CMF^{b}[\ypzc_{t+1}]&=\CMF^{b}[\ypzc_{t}] + \ln \frac{\popN^{\ypzc}_{t+1}}{\popN^{\ypzc}_{t}},\\
\CMF^{\cs}[\ypzc_{t+1},\zpzc_{t+1}]&=\CMF^{\cs}[\ypzc_{t},\zpzc_{t}] + \ln \frac{\popN^{\ypzc,\zpzc}_{t+1}}{\popN^{\ypzc,\zpzc}_{t}},
\end{align*}
where $\popN^{\ypzc}_{t}$ and $\popN^{\ypzc,\zpzc}_{t}$ are obtained by solving \eqnref{eq:popdyn} recursively for the given $\{\ypzc_{t},\zpzc_{t}\}$.
Similarly, $\CMF_{0}[\ypzc_{t}]$ is recursively computed as
\begin{align*}
\CMF_{0}[\ypzc_{t+1}]=\CMF_{0}[\ypzc_{t}] + \left(\phi_{0} + \ln \frac{\ProbT_{E}^{F}(y_{t+1}|y_{t})}{q_{0}(y_{t+1})} \right).
\end{align*}

The point-wise directed information $i[\zpzc_{t} \to \ypzc_{t}]$ in \eqnref{eq:DFRcSens} and \eqnref{eq:IFRcSens} is recursively obtained as 
\begin{align*}
i[\zpzc_{t+1} \to \ypzc_{t+1}] = i[\zpzc_{t} \to \ypzc_{t}] + \ln \frac{\ProbT_{E}^{F}(z_{t+1}|y_{t+1})}{\ProbT_{E}^{B}(z_{t+1}|y_{t})},
\end{align*}
where $\ProbT_{E}^{B}(z_{t+1}|y_{t})=\sum_{y_{t+1}}\ProbT_{E}^{F}(z_{t+1}|y_{t+1})\ProbT_{E}^{F}(y_{t+1}|y_{t})$.

Note that, when a feedback relation exists between $y_{t}$ and $z_{t}$,  the computation of $\CMF_{0}[\ypzc_{t}]$ and $i[\zpzc_{t} \to \ypzc_{t}]$ becomes more demanding because $\ProbQ[\ypzc_{t}]$ has to be obtained by marginalizing $\zpzc_{t}$ in $\ProbQ[\ypzc_{t},\zpzc_{t}]$.

\section*{References}


\begin{thebibliography}{83}
\bibitem{comment1}%
  \com{While the difference between phenotypic and environmental states is only $1$, this does not affect our numerical verifications because our theory is proved for more general situations. 
  However, if we consider more environmental states, we need more samples to verify the fluctuation relations, numerically.  }
\end{thebibliography}
%

\end{document}